\documentclass{aa}  
\usepackage{graphicx}
\usepackage{txfonts}
\usepackage{url}
\usepackage{natbib} 
\bibpunct{(}{)}{;}{a}{}{,} 
\begin{document}

\defcitealias{kupu00}{KP00}
\defcitealias{Vink00}{VdKL00}
   \title{Evolution towards the critical limit and the origin of Be stars}
 \titlerunning{Origin of Be stars}

   \author{S. Ekstr\"om
                  \and G. Meynet
                  \and A. Maeder
                 \and F. Barblan
          }
 \authorrunning{Ekstr\"om et al.}
   \offprints{Sylvia.Ekstrom@obs.unige.ch}

   \institute{Geneva Observatory, University of Geneva, Maillettes 51, 1290 Sauverny, Switzerland}

   \date{Received; accepted}

 \abstract
{More and more evidence lead to consider classical Be stars as stars rotating close to the critical velocity. If so, then the question which naturally arises is the origin of this high surface velocity.} 
  {We determine which are the mechanisms accelerating the surface of single stars during the Main Sequence evolution. We study their dependence on the metallicity and derive the frequency of stars with different surface velocities in clusters of various ages and metallicities.}
  {We have computed 112 stellar models of four different initial masses between 3 and 60 $M_{\sun}$, at four different metallicities between 0 and 0.020, and with seven different values of the ratio $\Omega/\Omega_\mathrm{crit}$ between 0.1 and 0.99. For all the models, computations were performed until either the end of the Main Sequence evolution or the reaching of the critical limit.}
{The evolution of surface velocities during the Main Sequence lifetime results from an interplay between meridional circulation (bringing angular momentum to the surface) and mass loss by stellar winds (removing it). The dependence on metallicity of these two mechanisms plays a key role in determining for each metallicity, a limiting range of initial masses (spectral types) for stars able to reach or at least approach the critical limit. Present models predict a higher frequency of fast rotating stars in clusters with ages between 10 and 25 Myr. This is the range of ages where most of Be stars are observed.
To reproduce the observed frequencies of Be stars, it is necessary to assume first that the Be star phenomenon occurs already for stars with $\upsilon/\upsilon_\mathrm{crit}\ge 0.7$ and second, that the fraction of fast rotators on the Zero Age Main Sequence is higher at lower metallicities. Depending on the stage at which the star becomes a Be star, the star at this stage may present more or less important enrichments in nitrogen at the surface.}
{}


   \maketitle

\section{Introduction}

The aim of the present work is to study how the surface velocity varies as a function of time during the Main Sequence (MS) evolution for stars of various initial masses, metallicities and rotation. Although predictions of the evolution of the surface velocity can already be found in literature \citep[see][]{Hel00,MMV, MMXI}, the published results lack homogeneity in their physical ingredients and cover too narrow ranges of initial velocities to provide valuable answers to the following questions:
\begin{itemize}
\item When is very fast rotation encountered in the course of the evolution of massive single stars on the MS phase?
\item What is the percentage of stars rotating faster than a given value of $\upsilon_\mathrm{eq}/\upsilon_\mathrm{crit}$ in stellar clusters of various ages and metallicities?
\item What are the expected chemical enrichments  at the surface of these very fast rotating stars? 
\item What are the initial conditions required for single stars to reach the critical velocity limit (defined as the rotational velocity such that the centrifugal acceleration exactly balances the gravity at the equator)?
\item Do these conditions depend on the mass and metallicity and if yes how?
\end{itemize}

To know which are the stars which can encounter the critical limit during the MS evolution is interesting in at least two astrophysical contexts: first, the study of Be stars, second, the very metal-poor stars representative of the stellar generations in the early Universe. Let us briefly describe the link with these two research areas.

\subsection{Be stars and the critical limit}

It is now widely accepted that the ``Be phenomenon'' is linked to fast rotation. Indeed, classical Be stars are B-type stars that exhibit line emission over the photospheric spectrum. This emission probably arises from an outwarding equatorial disk \citep{PR03} whose formation is (at least in part) due to the fast rotation of the star. If the star is rotating with $\upsilon_\mathrm{eq}/\upsilon_\mathrm{crit} \sim 0.95$, the additional velocity necessary to launch material into orbit is of the same order as the speed of sound in the outer layers \citep{Town04}. Therefore many instabilities, as {\it e.g.} non-radial pulsations, begin to be effective for orbital ejection \citep{Owocki}. Very interestingly, recent interferometric observations brought strong support to the idea that these stars are very fast rotators:
the oblateness of the Be star Achernar \citep{domi} suggests indeed that the star is rotating near the critical limit; \citet{Mei07} obtained for the first time direct evidence that the disk around the Be star $\alpha$ Arae is in Keplerian rotation. The wind geometry is compatible with a thin disk wind and a polar enhanced wind. These authors also found that $\alpha$ Arae is rotating very close to its critical rotation. Even the mystery behind the $\beta$ Ceph case seems to be elucidated: $\beta$ Ceph is a known Be star but also a known slow rotator, thus challenging the fast rotation scenario. However $\beta$ Ceph is a multiple system. \citet{Sch06} have shown that $\beta$ Ceph, the slow rotator, produces most of the light but is not responsible for the emission lines. The emission lines originate from a much fainter companion, the Be star, which they consider as probably fast rotating.

If the above observations do support the view that Be stars are at the critical limit, there are still some doubts that all Be stars are indeed at this limit. For instance \citet{Cra05}, on the basis of the observed distribution of the $\upsilon\sin i$ for Be stars, conclude that while late type Be stars could be at the critical limit, early Be type stars might rotate at only 40\% of their critical velocity. Let us note that the interpretation of the observations is stained with some theoretical uncertainties. A good example 
is the way the deduced velocities are affected by the fact that for a fast rotating star most of the radiative flux originates from the polar regions \citep{vZ24}. The low velocity regions, near the pole, will thus have a stronger impact on the spectrum than the dimmer faster regions near the equator. This implies that the rotational velocity of very fast rotators have been probably systematically underestimated \citep{coll04,Town04}. \citet{Cra05} has accounted for this effect in his work, but the problem is sufficiently complex that the results are uncertain.

Anyway, at the critical limit or not, Be stars represent wonderful laboratories for studying the effects of extreme rotation.

\subsection{Very metal poor stars and the critical limit}

The link between metal-free or very metal-poor stars and the question of the evolution of the surface velocity certainly deserves great attention, since it might deeply affect the evolution of these stars. Let us recall that at low metallicity, the radiation driven stellar winds are much weaker than at high metallicity, thus it is commonly accepted that mass loss for very metal-poor stars is very inefficient. This might not be true however when rotation is taken into account in the models for at least two reasons \citep[see][to get a more detailed discussion of these points]{NICIX}: first and somewhat paradoxically, since the stars are losing very little amount of mass by radiatively driven stellar winds, they have more chance to lose large amount of mass by rotational ejection during the MS evolution. Indeed, because of the low line driven winds, they will not be able to lose their angular momentum. They are thus more prone to reach the critical limit during their evolution \citep{MMVII,MEM06}. At the critical limit, as in the Be case, stellar matter is launched into an equatorial disk which will eventually dissipate by radiative effects. In that case the material will be lost by the star. If correct, such an evolution would thus promote mass loss through a mechanical wind triggered by rotation. This ``mechanical wind'', as stressed above, appears to be more important in metal-poor regions where the radiation driven stellar winds are expected to be very low. Second, rotational mixing enhances the surface CNO content which will lead to some increase of the outer layers opacities. This may revive the line driven winds and lead to strong mass loss.

Such mass loss, either triggered by rotational ejection or through the enhancement of the CNO content of the outer layers, have (if realized) important consequences on the evolution of the first stellar generations as well as on their nucleosynthesis \citep{EkTar,Hi07,De06}. Also such mass loss may be important in the process leading to the formation of a collapsar which is the most favoured model for the progenitors of the long Gamma Ray Bursts \citep{W93,YL05,WH06,MMGRB07}.

In the present work, we use rotating stellar models to address the questions of:
\begin{itemize}
\item the possible origin of Be stars from a purely theoretical point of view: can models reproduce the way the number of Be stars varies with the metallicity or with the age? Is their number sufficient to account for the observed number of Be stars (supposing that Be stars are star rotating with values of $\upsilon_\mathrm{eq}/\upsilon_\mathrm{crit}$ superior to a given value)?
\item the behaviour of rotating stars at very low or zero metallicity: how does the lack of metals influence the evolution of the equatorial velocity?
\end{itemize}
Of course the answers given to these questions will depend strongly on the stellar models used. However we think that the exercise deserves to be undertaken for the following reason: the stellar models we use were computed with physical ingredients that made predictions consistent with many observational features of massive stars \citep[see][and the references therein on the series of papers by the Geneva group on rotating models]{MMXI} and thus it is interesting to explore in more detail what these models say about fast rotating stars. For very fast rotation, the present models probably need still to be improved by the incorporation of additional physics (e.g. pulsation), and we hope that the results obtained here will guide us towards improvements in the future.

The paper is organized as follows: in Sect.~\ref{inpphys} the physical ingredients of the models are presented. We also recall some basic relations from the Roche model useful for discussing some properties of our models on the ZAMS. In Sect.~\ref{zams}, we present various characteristics of our models on the ZAMS as a function of the initial mass, velocity and metallicity. The evolution of the models on the MS is presented in Sect.~\ref{evolms}. The question of the initial conditions required to reach a given value of $\upsilon_\mathrm{eq}/\upsilon_\mathrm{crit}$ during the MS phase is the subject of Sect.~\ref{bul}. The frequency of stars with velocities exceeding a given value of $\upsilon_\mathrm{eq}/\upsilon_\mathrm{crit}$ is presented in Sect~\ref{freqbe}. The sensitivity of the predicted frequencies on various hypotheses made in this work are discussed in Sect.~\ref{discuss}.
Possible links with observational features of Be stars will be presented in Sect~\ref{becomp}. Conclusions and future perspectives are the subject of Sect.~\ref{conclu}.

Let us mention that electronic tables are available for all our ZAMS and evolutionary models at \url{http://obswww.unige.ch/Recherche/evol/Critical-limit-and-Be-stars}

\section{Physics of the models \label{inpphys}}

\subsection{Input physics}
Since the detailed description of the physics of the models has already been presented
in detail in previous papers (see references below), we shall keep here the discussion very short.
The physics of rotation (transport of the angular momentum and of the chemical species) are treated as in \citet{MMXI}. These models are based on the theory 
of the transport mechanisms induced by rotation proposed by \citet{Za92} and further
complemented by the works of \citet{TZ97} and \citet{MZ98}.

We have computed the evolution of five different masses, $M=$ 1, 3, 9, 20, 60 $M_{\sun}$, at four different metallicities, $Z=$ 0.020, 0.002, 0.00001, 0. For each combination of $M-Z$, we have computed models with seven different initial rotational rates, $\omega=\Omega/\Omega_\mathrm{crit}=$ 0.10, 0.30, 0.50, 0.70, 0.80, 0.90, 0.99. The 3 to 60 $M_{\sun}$ models have been evolved until either the beginning of core helium burning or the reaching of break-up limit. For the 1 $M_{\sun}$ models, only ZAMS models are presented.

The initial composition of the models for the metallicities $Z=0.020$, 0.002, 0.00001 and 0 are respectively $X=0.7050$, 0.7545, 0.7599 and 0.76 and $Y=0.275$, 0.2435, 0.24002 and 0.24. For the heavy elements, we have used the same mixture as the one used in the OPAL opacity tables: solar mixture of Grevesse \& Noels (1993) for $Z=0.020$ and 0.002 and $\alpha$-enhanced mixture
of Weiss (1995) for $Z=0.00001$\footnote{see \url{http://www-phys.llnl.gov/Research/OPAL/opal.html} for exact references}. The nuclear reaction rates are from the NACRE database \footnote{ see
\url{http://pntpm.ulb.ac.be/Nacre/barre_database.htm}}.

The convection is treated according to the Schwarzschild criterion. The size of the convective core is increased by a moderate overshooting of 0.1 H$_\mathrm{p}$.

The 3 $M_{\sun}$ models were computed without mass loss. The 9 $M_{\sun}$ models were computed with \citet{dJ88} prescription. For the higher mass models we have used the prescription proposed by \citealt{kupu00} \citepalias[hereafter][]{kupu00}, and \citet{dJ88} in the temperature domain not covered by the first prescription. The effects of rotation on the mass loss rates are taken into account as explained in \citet{mm6}.

\subsection{The Roche model \label{roche}}

In all the derivations, we shall use the Roche model, {\it i.e.} we approximate the gravitational potential by $GM_r/r$ where $M_r$ is the mass inside the equipotential with a mean radius $r$. The radius $r$ which labels each equipotential is defined by $r=\left( V_r/(4/3 \pi)\right )^{1/3}$ where $V_r$ is the volume (deformed by rotation) inside the equipotential considered.\footnote{ZAMS models are supposed to rotate like solid bodies, the centrifugal force is conservative and it is possible to define equipotential surfaces. This is no longer the case for shellular rotation. In that case however, the shape of the isobaric surfaces are given by the same expression as the one giving the shape of the equipotentials for solid body rotation provided some changes of variables are performed \citep[see][]{Me97}.} 

In the frame of the Roche model and for a conservative rotation law ({\it i.e.} such that the centrifugal force can be derived from a potential), the shape of a meridian at the surface of an equipotential is given by couples of $r$ and $\theta$ values satisfying the following equation:
\begin{equation}
{GM_{r} \over r}+{1 \over 2}\Omega_{r}^2 r^2 \sin^2 \theta={GM_{r} \over r_\mathrm{p}},
\label{eq1}
\end{equation}
where $r$ is the radius at colatitude $\theta$, $\Omega_{r}$ the angular velocity at that radius, $M_{r}$ the mass inside the considered equipotential surface and $r_\mathrm{p}$, the radius at $\theta=0$. Thus the shape of a given equipotential is determined by three parameters $M_{r}$, $\Omega_{r}$ and $r_\mathrm{p}$. The first two $M_{r}$ and $\Omega_{r}$ are independent variables. The third one is derived from the first two and the equations of stellar structure. Setting $x=\left({GM_{r} \over \Omega_{r}^2}  \right)^{-1/3} r$, one can write Eq.~(\ref{eq1}) \citep[see][]{KippTh70}
\begin{equation}
{1 \over x}+{1 \over 2}x^2 \sin^2 \theta={1 \over x_\mathrm{p}}.
\label{eq1b}
\end{equation}
With this change of variable, the shape of an equipotential is uniquely determined by only one parameter $x_\mathrm{p}$.

With $M$ the total mass of the star and $\Omega$ the angular velocity at the surface, Eqs.~(\ref{eq1}) or (\ref{eq1b}) give the shape of a meridian line at the stellar surface. Setting
\begin{equation}
f={R_\mathrm{e} \over R_\mathrm{p}},
\label{eq2}
\end{equation}
where $R_\mathrm{e}$ is the equatorial radius, one easily obtains from Eq.~(\ref{eq1}) that
\begin{equation}
R_\mathrm{p}=\left({GM \over \Omega^2} \right)^{1/3} \left({2 (f-1)\over f^3}
\right)^{1/3}=\left({GM \over \Omega^2} \right)^{1/3} x_\mathrm{p}.
\label{eq3}
\end{equation}
The above equation relates the inverse of the oblateness $f$ to $R_\mathrm{p}$. 

As recalled above, the critical angular velocity, $\Omega_\mathrm{crit}$, corresponds to the angular velocity at the equator of the star such that the centrifugal force exactly balances the gravity. The classical critical angular velocity or the $\Omega$-limit (to distinguish it from the $\Omega\Gamma$-limit as defined in \citealt{mm6}) in the frame of the Roche model is given by 
\begin{equation}
\Omega_\mathrm{crit}=\left({2 \over 3}\right)^{3 \over 2}\left({GM \over R^3_\mathrm{p,crit}}\right)^{1 \over 2},
\label{eq4}
\end{equation}
where $R_\mathrm{p,crit}$ is the polar radius when the surface rotates with the critical velocity. The critical velocity is given by
\begin{equation}
\upsilon_\mathrm{crit}=\left({2 \over 3}{GM \over R_\mathrm{p,crit}}\right)^{1 \over 2}.
\label{eq4b}
\end{equation}
Replacing $\Omega$ in Eq.~(\ref{eq3}) by $(\Omega/\Omega_\mathrm{crit}) \cdot \Omega_\mathrm{crit}$ and using Eq.~(\ref{eq4}), one obtains a relation between $R_\mathrm{p}$, $f$ and $R_\mathrm{p,crit}$,
\begin{equation}
R_\mathrm{p}={3 \over 2} R_\mathrm{p,crit} \left({\Omega_\mathrm{crit} \over \Omega}\right)^{2/3} \left({2 (f-1)\over f^3} \right)^{1/3},
\label{eq5}
\end{equation}
Thus one has
\begin{equation}
{\Omega \over \Omega_\mathrm{crit}}=\left({3 \over 2}\right)^{3/2} \left({R_\mathrm{p,crit} \over R_\mathrm{p}}\right)^{3/2} \left({2 (f-1)\over f^3} \right)^{1/2}.
\label{eq6}
\end{equation}
With a good approximation (see below) one has that $R_\mathrm{p,crit}/ R_\mathrm{p} \simeq 1$ and therefore
\begin{equation}
{\Omega \over \Omega_\mathrm{crit}} \simeq \left({3 \over 2}\right)^{3/2} \left({2 (f-1)\over f^3} \right)^{1/2}.
\label{eq7}
\end{equation}
 \begin{figure}
  \resizebox{\hsize}{!}{\includegraphics[angle=0]{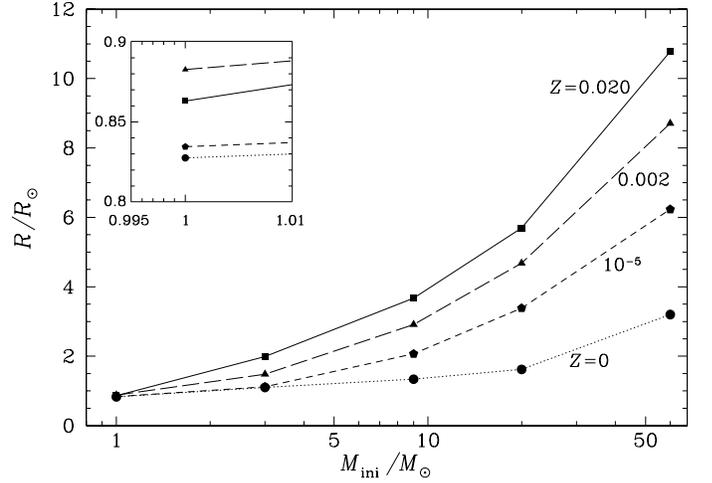}}
      \caption{Variations of the radius as a function of initial mass, for
      various metallicities for non-rotating stars. The inset zoom shows the small difference of radii in the case of the 1 $M_{\sun}$ models.}
       \label{rayons}
  \end{figure}

This equation is quite useful since it allows the determination of $\Omega_\mathrm{crit}$ from quantities obtained with a model computed for any $\Omega < \Omega_\mathrm{crit}$, however the closer to the critical value, the better. It is not the case of Eq.~(\ref{eq5}) which involves $R_\mathrm{p,crit}$ whose knowledge can only be obtained by computing models at the critical limit. Below we shall see to what extent the use of expression (\ref{eq7}) provides a good approximation of $\Omega_\mathrm{crit}$.

Setting $\upsilon$ the velocity at the equator, one has that
\begin{equation}
{\upsilon \over \upsilon_\mathrm{crit}}={\Omega R_\mathrm{e}\over\Omega_\mathrm{c}R_\mathrm{e,crit}}
={\Omega \over \Omega_\mathrm{crit}}{R_\mathrm{e}\over R_\mathrm{p}}{R_\mathrm{p}\over R_\mathrm{p,crit}}{R_\mathrm{p,crit}\over R_\mathrm{e,crit}},
\label{eq7bis}
\end{equation}
where $R_\mathrm{e,crit}$ is the equatorial radius when the surface rotates with the critical velocity.
Using Eq.~(\ref{eq5}) above, and the fact that in the Roche model $R_\mathrm{p,crit}/ R_\mathrm{e,crit}=2/3$,
one obtains
\begin{equation}
{\upsilon \over \upsilon_\mathrm{crit}}=
\left({\Omega \over \Omega_\mathrm{crit}}2(f-1)\right)^{1/3}
\label{eq8}
\end{equation}

An interesting quantity is the ratio of the centrifugal acceleration $a_\mathrm{cen}$ to the gravity $g_\mathrm{e}$ at the equator
\begin{equation}
{a_\mathrm{cen} \over g_\mathrm{e}}={\Omega^2 R^3_\mathrm{e}\over G M}=
\left({\Omega \over \Omega_\mathrm{crit}}\right)^2 \left({2 \over 3}\right)^3 f^3
\left({R_\mathrm{p}\over R_\mathrm{p,crit}}\right)^3,
\label{eq9}
\end{equation}
where we have used Eq.~(\ref{eq4}) and divided/multiplied by $R_\mathrm{p,crit}^3$. Replacing $R_\mathrm{p,crit}/R_\mathrm{p}$ by its expression deduced from Eq.~(\ref{eq5}), we obtain
\begin{equation}
{a_\mathrm{cen} \over g_\mathrm{e}}=2(f-1).
\label{eq10}
\end{equation}
We can check that at the critical limit, when $f=3/2$, then $a_\mathrm{cen}=g_\mathrm{e}$.

\section{Basic parameters on the Zero Age Main Sequence \label{zams}}
We consider that a star has arrived on the ZAMS, when a mass fraction of 0.003 of hydrogen has been transformed into helium at the centre. At this stage, the star is supposed to have a solid body rotation.
We present here the stellar radii, the total luminosity, the variation with the colatitude of the effective temperature and the moment of inertia of the models. We will then show the relations between the initial rotation and the angular momentum content, $\upsilon_\mathrm{ini}/\upsilon_\mathrm{crit}$ and $\Omega_\mathrm{ini}/\Omega_\mathrm{crit}$. 

\subsection{Radii, luminosities and effective temperatures \label{rlt}}

Considering first the case without rotation, we can see in Fig.~\ref{rayons} that there is a relation between radius and mass of the form $R\sim M^{0.6}$ at standard metallicity. This relation becomes shallower with decreasing $Z$, down to $R\sim M^{0.3}$ at $Z=0$. This comes mainly from two causes: 1) first, at lower metallicity, the opacities are lower; 2) the CNO elements are less abundant, or even absent in metal-free stars. In order to compensate for the energy radiated by the surface when smaller amounts of CNO elements are present, the central regions must extract energy from nuclear reactions at a higher temperature regime. To reach this regime, the star contracts more during the pre-MS phase and is thus more compact on the ZAMS. For example, in the case of the 60 $M_{\sun}$ models, the radius at $Z=0$ is reduced by a factor 3.36 in comparison with the radius at standard metallicity. The 1 $M_{\sun}$ models are running only on $pp$-chains and show very small differences in their radii: only about 5\% between $Z=0.020$ and $Z=0$ (see the inset zoom in Fig.~\ref{rayons}).
  \begin{figure}
  \resizebox{\hsize}{!}{\includegraphics[angle=-90]{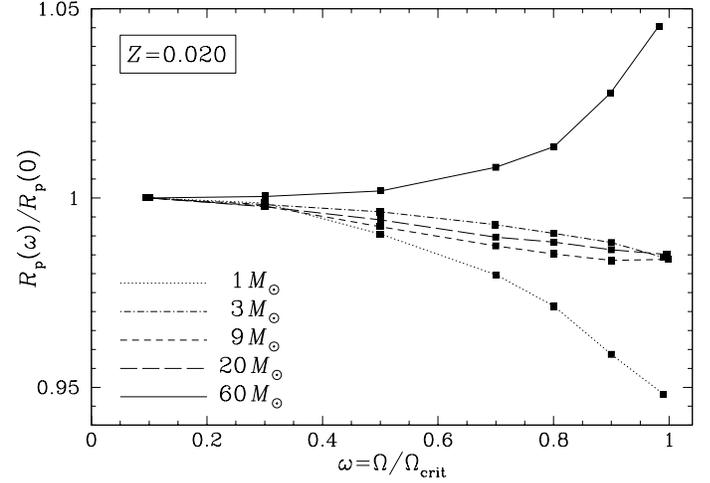}}
      \caption{Variations of the polar radius as a function of the ratio
      $\omega=\Omega/\Omega_\mathrm{crit}$, normalized to the non-rotating value,
      for various masses at standard metallicity.}
       \label{rprpo}
  \end{figure}  

In Fig.~\ref{rprpo}, we present the variation of the polar radius $R_\mathrm{p}$ as a function of $\omega=\Omega/\Omega_\mathrm{crit}$, normalized to its non-rotating value, for the standard metallicity models. We see that increasing $\omega$ tends to decrease the radius in the case of the 1 to 20 $M_{\sun}$, a trend which can  be explained as follows: rotation plays the role of a sustaining force in the star. Thus, the central temperature and density in rotating stars are lower, decreasing subsequently the luminosity and the radius. The change of the polar radius for the 1 $M_{\sun}$ model amounts to about 5\%. In the case of the 3, 9 and 20 $M_{\sun}$\ models, the assumption of $R_\mathrm{p,crit}/R_\mathrm{p}=1$ (Sect.~\ref{roche}) is absolutely valid, since the decrease amounts only to 1.5\%. Why are nearly completely radiative stars as the 1 $M_{\sun}$ stellar models more deformed for a given value of $\omega$ than stars with a significant convective core? It comes from the fact that the density gradient in convective zones are shallower than in radiative ones, allowing more mass to be contained in a given volume. This strengthens gravity and the resistance to deformation.

\begin{figure}
  \resizebox{\hsize}{!}{\includegraphics[angle=-90]{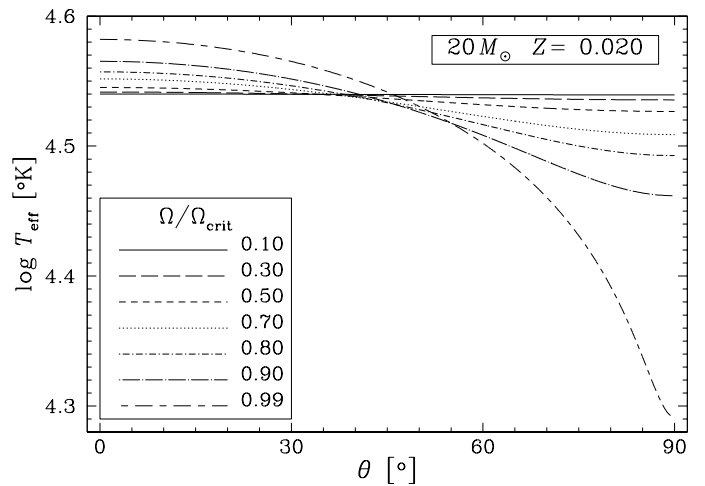}}
      \caption{Variations of the effective temperature $T_\mathrm{eff}$ as a
      function of colatitudinal angle $\theta$, for the various values of
      rotational rates, in the 20 $M_{\sun}$ model at standard metallicity.}
       \label{teffcolat}
  \end{figure}

For the 60 $M_{\sun}$ models, the change of the polar radius is of the same amplitude as for the 1 $M_{\sun}$ model but in the opposite direction, showing an increase of about 4.5\%. How could be explained such a difference of behaviour between the lowest and highest mass models? In the high mass star range, radiation pressure contributes more to the total pressure. It is the highest in the polar regions where the effective gravity (gravity decreased by the centrifugal acceleration) is the highest (von Zeipel theorem). A look at Fig.~\ref{teffcolat} shows that $T_\mathrm{eff}$(pole) is higher when $\omega$ is higher (10\% between $\omega=0.1$ and $\omega=0.99$), and therefore $P_\mathrm{rad}\sim T^4$ is stronger. This tends to inflate the polar radius in massive stars.
 \begin{figure}
  \resizebox{\hsize}{!}{\includegraphics[angle=-90]{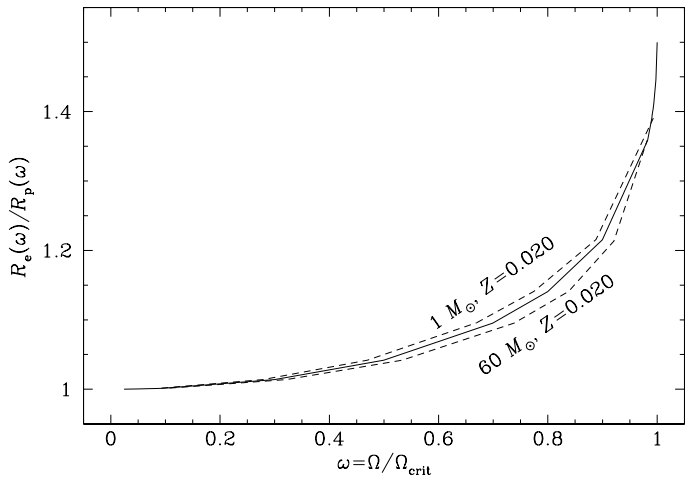}}
      \caption{Variation of the ratio $R_\mathrm{e}(\omega)/R_\mathrm{p}(\omega)$ (equatorial over
      polar radius) as a function of $\omega=\Omega/\Omega_\mathrm{c}$. The
      continuous line shows the relation given by Eq.~(\ref{eq7}), assuming
      $R_\mathrm{p,crit}/R_\mathrm{p}=1$. The dashed lines show the relations for the
      $Z=0.020$ models with 1 and 60 $M_{\sun}$ using Eq.~(\ref{eq6}).}
       \label{rerp}
  \end{figure}

Rotation deforms the star, the ratio $R_\mathrm{e}/R_\mathrm{p}$ increasing with $\omega$ as shown in Fig.~\ref{rerp}. We have plotted the simplified relation obtained by Eq.~(\ref{eq7}), assuming
$R_\mathrm{p,crit}/R_\mathrm{p}=1$ (continuous line), as well as the relation obtained using Eq.~(\ref{eq6}) applied to two different masses (1 and 60 $M_{\sun}$) at $Z=0.02$. In this case, we observe the direct effect of the behaviour seen in Fig.~\ref{rprpo}. For the 1 $M_{\sun}$, the decrease of $R_\mathrm{p}$ with increasing $\omega$ makes $f$ higher than the continuous line, joining it only at the critical limit since, in the Roche model frame, $R_\mathrm{e,crit}/R_\mathrm{p,crit}$ is always equal to 1.5, while the inverse behaviour of the 60 $M_{\sun}$ leads to the inverse result. Let us note also that in all cases, the equatorial radius becomes longer than 1.1 times the polar one only for $\omega \geq 0.75$.
  \begin{figure}
  \resizebox{\hsize}{!}{\includegraphics[angle=-90]{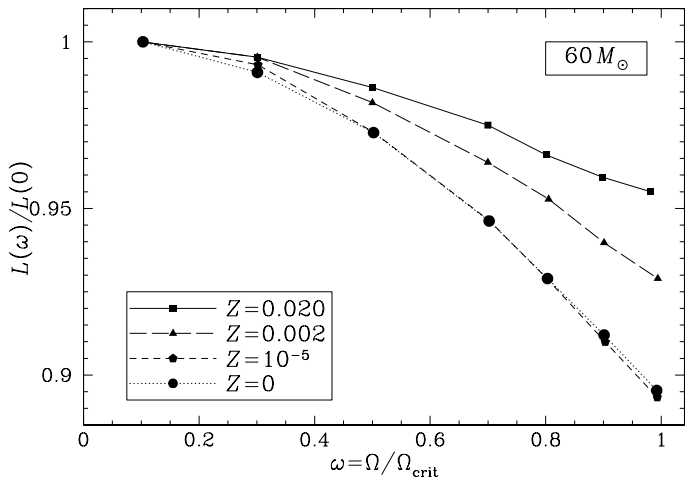}}
      \caption{Variations of the total luminosity as a function of $\omega$,
      normalized to the non-rotating value, for 60 $M_{\sun}$ at various
      metallicities.}
       \label{lum60}
  \end{figure}

In Fig.~\ref{lum60}, we show the behaviour of the luminosity $L(\omega)$ of our 60 $M_{\sun}$ models as a function of $\omega$ for the different metallicities considered. The value of the luminosity is normalized to its non-rotating value. One sees that the luminosity decreases by at most 5 to 10\% when rotation increases up to the critical limit. As mentioned above, it is due to the fact that the centrifugal acceleration helps in sustaining the star against the gravity and allows the luminosity to take a value corresponding to a non-rotating lower initial mass star. We can see that the effects of rotation are stronger at lower metallicity. This is related to the fact that low metallicity stars are more compact. The regions having a temperature high enough to undergo nuclear burning are more extended in the star and thus overlap regions that are more affected by rotation.
  \begin{figure}
  \resizebox{\hsize}{!}{\includegraphics[angle=-90]{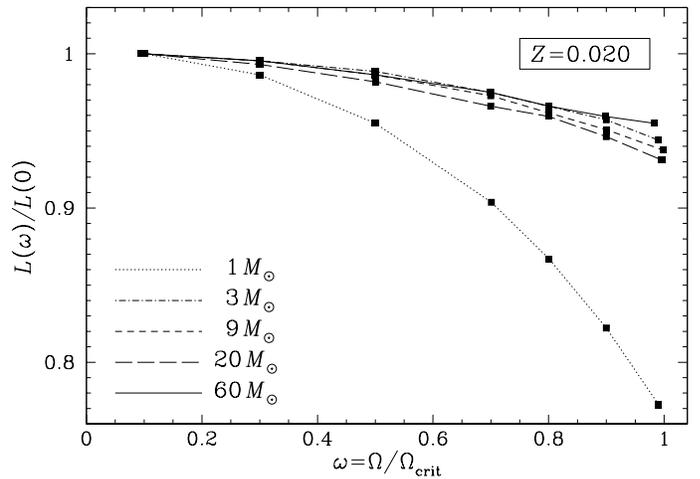}}
      \caption{Variations of the total luminosity as a function of $\omega$,
      normalized to the non-rotating value, for various masses at standard
      metallicity.}
       \label{llo}
  \end{figure}

Fig.~\ref{llo} is the same as Fig.~\ref{lum60}, but this time at fixed metallicity ($Z=0.020$) and varying the mass. The 1 $M_{\sun}$ is the most affected by rotation, with a decrease of 22.5\% of its luminosity at break-up. The higher masses show a decrease of 4.5 to 7\% only. This again is related to the fact that in the 1 $M_{\sun}$ model, the energy production through $pp$-chains is more extended in the interior of the star. As above, a non-negligible part of the luminosity is produced in regions that may suffer from the effects of rotation.

Figure \ref{teffcolat} shows the variation with colatitude of the effective temperature at the surface of a 20 $M_{\sun}$ star for various rotation rates at standard metallicity. The ratio ($T_\mathrm{eff}$(pole)-$T_\mathrm{eff}$(equator))/$T_\mathrm{eff}$(equator) becomes superior to 10\% only for $\omega > 0.7$. Near break-up, the effective temperature of the polar region is about a factor two higher than that of the equatorial one. We may compare this result with the recent work of \citet{ELR07} who have studied the structure and dynamics of rapidly rotating stellar models with a two-dimension code. They have computed a model with $\omega=0.82$ for which the ratio ($T_\mathrm{eff}$(pole)-$T_\mathrm{eff}$(equator))/$T_\mathrm{eff}$(equator) is 14\%. Our closer model ($\omega=0.80$) shows a slightly higher ratio of 16\% though we should expect a smaller value due to the lower $\omega$. Let us note that the 2D simulation has been carried out within a spherical box which, according to the authors, may soften the latitudinal variations. It would be interesting to check how much of the discrepancy remains would the 2D computation be done without such a container.

\subsection{Inertia, angular momenta and rotational energies}
  \begin{figure}
  \resizebox{\hsize}{!}{\includegraphics[angle=-90]{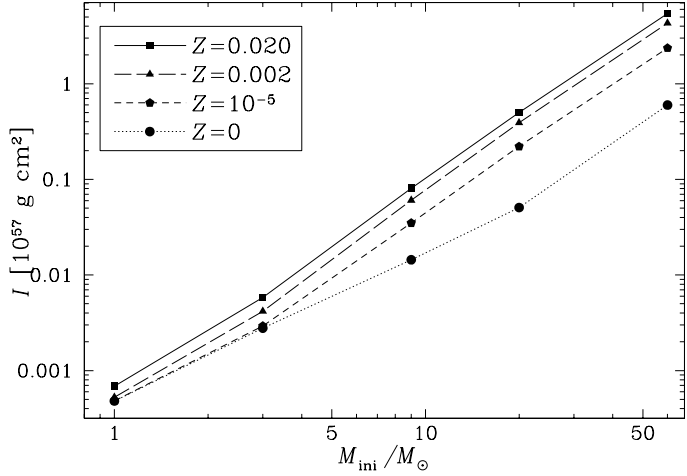}}
      \caption{Variations of the moment of inertia as a function of initial mass
      for various metallicities (models with $\Omega/\Omega_\mathrm{crit}=0.1$).}
       \label{inertm}
  \end{figure}

Fig.~\ref{inertm} presents the variations of the moment of inertia with the initial mass of the models. These variations show almost no dependence on $\omega$, so only the case of $\omega=0.1$ is plotted. The moments of inertia have
been obtained by summing the contributions of each shell $i$ of the model: $I=\sum_{i} \ (2/3)\, r_{i}^2\, {\rm d}\,m_{i}$, where $r_{i}$ is the mean radius of the shell, \emph{i.e.} the radius in $P_2 (\cos \theta) =0$. Thus, $r_{i}$ accounts for the rotational distortion up to the second order Legendre polynomial, and the above way of calculating the moment of inertia is correct at this level of approximation. Let us mention that rotation deforms only the outermost layers of the model, which contain very little mass. This is why the moment of inertia does not change much with or without rotation.

We note that, at a given metallicity, the moment of inertia $I$ increases with the initial mass. The increase amounts to about 4 orders of magnitudes between 1 and 60 $M_{\sun}$. As can be seen on Fig.~\ref{inertm}, the relation between $I$ and $M$ is almost linear in logarithmic scale, with a slope of 2.2 for the non-zero metallicities and of 1.7 for $Z=0$. The difference of the slopes can be linked to the difference in the way the radii increase with increasing mass. From the mass-radius relations obtained in Sect.~\ref{rlt} ($R\sim M^{0.6}$ at standard metallicity and $R\sim M^{0.3}$ at $Z=0$) we indeed expect a relation $I\sim M^{2.2}$ and $I\sim M^{1.6}$ at $Z=0.02$ and 0 respectively. For a given initial mass, the moments of inertia are smaller at lower metallicities, as expected due to smaller radii at lower $Z$.
  \begin{figure}
  \resizebox{\hsize}{!}{\includegraphics[angle=-90]{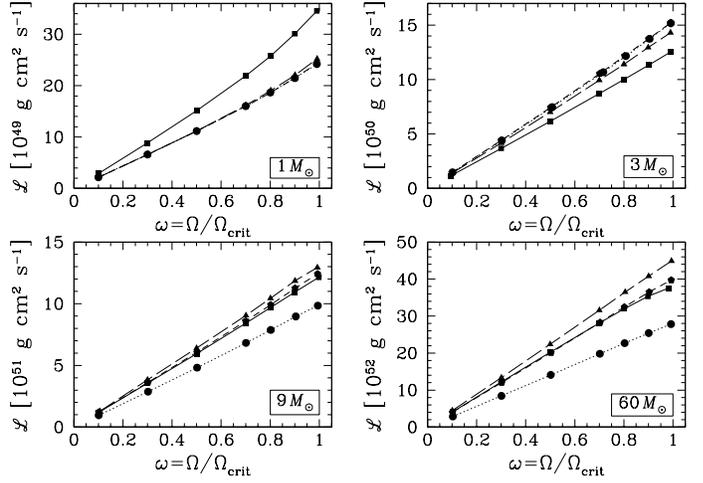}}
      \caption{Variations of the total angular momentum content of the models on
      the ZAMS as a function of initial rotation rate $\omega = \Omega/\Omega_\mathrm{crit}$
      for various metallicities, plotted according to the initial mass of the
      models. The coding for $Z$ is the same as in Fig.~\ref{inertm}}
       \label{ltot}
  \end{figure}
  \begin{figure}
  \resizebox{\hsize}{!}{\includegraphics[angle=-90]{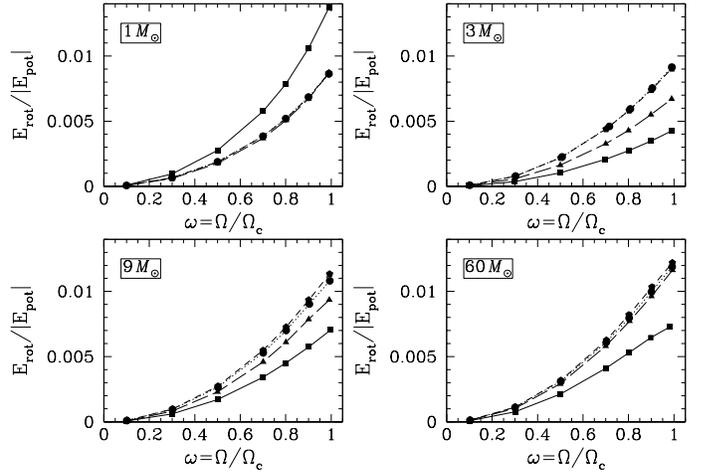}}
      \caption{Variations of the ratio between rotational and potential energy
      as a function of the ratio $\omega=\Omega/\Omega_\mathrm{crit}$ for various
      metallicities, plotted according to the initial mass of the model. The coding
      for $Z$ is the same as in Fig.~\ref{inertm}}
       \label{rotpot}
  \end{figure}

The variation of the total angular momentum content ($\mathcal{L}=\sum_{i} \ I_{i}\,\Omega_{i}$) as a function of  $\omega$ is shown in Fig.~\ref{ltot}. Note that the highest point on the right of the figures gives an estimate of the maximum angular momentum content on the ZAMS (for solid body rotation). Since the moment of inertia of the star does not depend much on the rotation velocity, and since on the ZAMS, we suppose solid body rotation, one obtains quasi linear relations between $\mathcal{L}$ and $\omega$. We note that the metallicity dependence remains modest at least in the range of metallicities between 0.00001 and 0.020. In the 9 to 60 $M_{\sun}$ range, the value of $\omega$ being kept constant, there is at first a slight increase of the angular momentum content when decreasing the metallicity (compare the curve with the black squares for the $Z=0.020$ models with the curve with the triangles corresponding to 0.002 models). Then for still lower metallicities, the angular momentum decreases with the metallicity (the curve for the $Z=0.00001$ models overlaps the $Z=0.020$ models) reaching its lower values for the metal-free stellar models.

In the mass range between 9 and 20 $M_{\sun}$, stars having the same angular momentum content on the ZAMS would have very similar value of $\omega$ whatever their metallicity between 0.00001 and 0.02. Only if the metallicity is zero, would the value of $\omega$ corresponding to the same angular momentum content be much higher. As a numerical example, a 20 $M_{\sun}$ on the ZAMS with $\mathcal{L}$= 0.3$\cdot$10$^{53}$ cm$^2$ g sec$^{-1}$ has a ratio $\omega=~0.5$ when $Z$ is comprised between 0.00001 and 0.020. This angular momentum content corresponds to $\omega=~0.8$ when $Z=0$.
  \begin{figure}
  \resizebox{\hsize}{!}{\includegraphics[angle=-90]{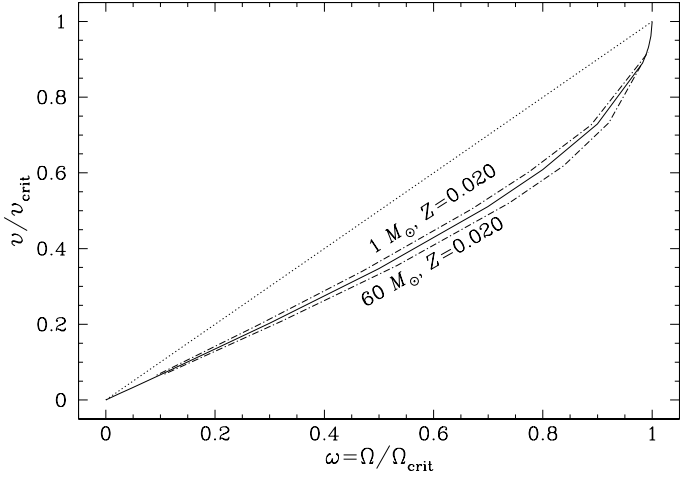}}
      \caption{Relation between $\upsilon/\upsilon_\mathrm{crit}$ and
      $\omega= \Omega/\Omega_\mathrm{crit}$ obtained in the frame of the Roche model. The
      continuous line is obtained assuming $R_\mathrm{p,crit}/R_\mathrm{p}=1$ (see text
      and Eqs.~(\ref{eq7}) and (\ref{eq8})). The dot-dashed lines show the relations
      for the $Z=0.020$ models with 1 and 60 $M_{\sun}$ using Eq.~(\ref{eq6}). The
      dotted line is the line of slope 1.}
       \label{ocvc}
  \end{figure}
  \begin{figure}
  \resizebox{\hsize}{!}{\includegraphics[angle=-90]{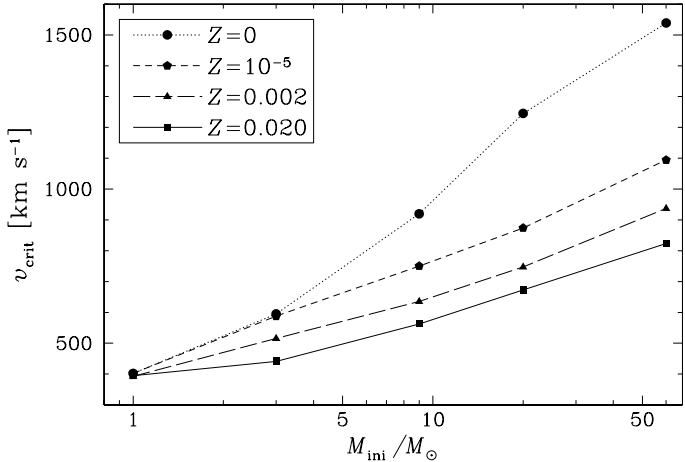}}
      \caption{Variations of the critical equatorial velocity on the ZAMS as a
      function of the initial mass, for various metallicities.}
       \label{vcrit}
  \end{figure} 

The rotational kinetic energy ($E_\mathrm{rot}=\sum_{i} \ (1/2)\, I_{i}\, \Omega_{i}^2$) expressed as a fraction of the binding energy ($E_\mathrm{|pot|}=-G \sum_{i} \ (M_{i}/r_{i})\, {\rm d}\,m_{i}$) is shown in Fig.~\ref{rotpot}. The rotational energy amounts to at most a percent of the binding energy. This is consistent with the well known fact that the effects of the centrifugal acceleration remains quite modest on the hydrostatic structure of the stellar interior, even when the surface rotates with a velocity near the critical one.  For a 60 $M_{\sun}$ stellar model at $Z=0.020$ the binding energy is $1.8 \cdot 10^{51}$ ergs, its $Z=0$ counterparts has a binding energy equal to $5.3 \cdot 10^{51}$ ergs, {\it i.e.} about a factor 3 higher. Let us recall for comparison that the binding energy of a neutron star is of the order of 10$^{53}$ ergs, about two orders of magnitude greater.    

\subsection{Relations between $\Omega/\Omega_\mathrm{crit}$ and $\upsilon/\upsilon_\mathrm{crit}$} 
 \begin{figure}
  \resizebox{\hsize}{!}{\includegraphics[angle=-90]{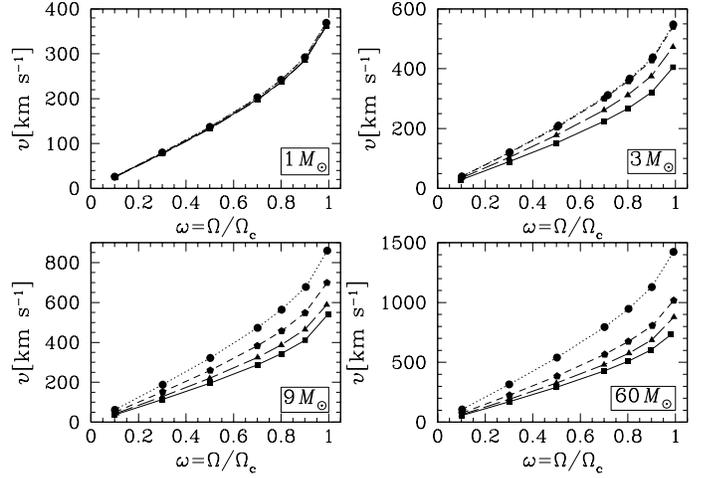}}
      \caption{Variations of the surface equatorial velocity on the ZAMS as a
      function of initial rotation rate $\omega = \Omega/\Omega_\mathrm{crit}$ for various
      metallicities and masses. The coding for $Z$ is the same as in Fig.~\ref{vcrit}.}
       \label{vsurf}
  \end{figure}  
%
The relations between $\upsilon/\upsilon_\mathrm{crit}$ and $\Omega/\Omega_\mathrm{crit}$ obtained in the frame of the Roche model (see Eq.~(\ref{eq8})) for the 1 and 60 $M_{\sun}$ stellar models at $Z=0.02$ are shown in Fig.~\ref{ocvc}. In case we suppose that the polar radius remains constant ($R_\mathrm{p,crit}/R_\mathrm{p}=1$), then  Eq.~(\ref{eq7}) can be used and one obtains a unique relation between $\Omega/\Omega_\mathrm{crit}$ and $\upsilon/\upsilon_\mathrm{crit}$, independent of the mass, metallicity and evolutionary stage considered (continuous line). One sees that the values of $\upsilon/\upsilon_\mathrm{crit}$ are smaller than that of $\Omega/\Omega_\mathrm{crit}$ by at most $\sim$25\%. At the two extremes the ratios are of course equal.  
\begin{table}[t]
\caption{Angular velocity, rotational period, polar radius and equatorial
velocity at the critical limit on the ZAMS (solid body rotation).} \label{rotcar}
\begin{center}\scriptsize
\begin{tabular}{llcrrr}
\hline \hline
    &      &                  &        &              &                    \\
$M$ & $Z$  & $\Omega_\mathrm{crit}$ & P$_\mathrm{crit}$ & $R_\mathrm{p,crit}$ &
$\upsilon_\mathrm{crit}$ \\
$M_{\sun}$ &   & s$^{-1}$ & hours & $R_{\sun}$ & km s$^{-1}$ \\
    &      &                  &        &              &                    \\
\hline     
    &      &                  &        &              &                    \\      
     1  &   0        &  4.9E-04  &    3.6  &    0.8  &   400 \\
     1  &   $10^{-5}$  &  4.9E-04  &    3.6  &    0.8  &   400 \\
     1  &   0.002    &  4.6E-04  &    3.8  &    0.8  &   390 \\
     1  &   0.020     &  4.6E-04  &    3.8  &    0.8  &   395 \\
    &      &                  &        &              &                    \\
     3  &   0        &  5.3E-04  &    3.3  &    1.1  &   595 \\
     3  &   $10^{-5}$  &  5.1E-04  &    3.4  &    1.1  &   585 \\
     3  &   0.002    &  3.4E-04  &    5.1  &    1.4  &   515 \\
     3  &   0.020    &  2.2E-04  &    8.1  &    2.0  &   440 \\
            &            &            &          &          &       \\     
     9  &   0        &  6.5E-04  &    2.7  &    1.4  &   920 \\
     9  &   $10^{-5}$  &  3.5E-04  &    4.9  &    2.0  &   750 \\
     9  &   0.002    &  2.2E-04  &    8.1  &    2.8  &   635 \\
     9  &   0.020     &  1.5E-04  &   11.7  &    3.6  &   560 \\
            &            &            &          &          &       \\     
    20  &   0        &  7.3E-04  &    2.4  &    1.6  &  1245 \\
    20  &   $10^{-5}$  &  2.5E-04  &    7.0  &    3.3  &   870 \\
    20  &   0.002    &  1.6E-04  &   11.1  &    4.6  &   745 \\
    20  &   0.020     &  1.2E-04  &   15.2  &    5.6  &   670 \\
            &            &            &          &          &       \\    
    60  &   0        &  4.6E-04  &    3.8  &    3.2  &  1540 \\
    60  &   $10^{-5}$  &  1.7E-04  &   10.6  &    6.4  &  1095 \\
    60  &   0.002    &  1.0E-04  &   16.9  &    8.7  &   935 \\
    60  &   0.020     &  7.0E-05  &   24.9  &   11.3  &   820 \\
            &            &            &          &          &       \\
\hline
\end{tabular}
\end{center}
\end{table}
\begin{table}[t]
\caption{The first line indicates for each initial mass and metallicity,
the equatorial velocity in km s$^{-1}$ on the ZAMS. For the 3, 20 and 60 $M_{\sun}$ stellar
mass models the second line gives
the time-averaged value of the velocity over the adjustment period from solid to non-solid rotation
at the beginning of the MS phase (see Sect.~4.2).}
\label{veloc}
\begin{center}\scriptsize
\begin{tabular}{ccccccccc}
\hline \hline
    &           &      &       &       &       &       &        &        \\
M   &  $Z$      & \multicolumn{7}{c}{$\Omega/\Omega_\mathrm{crit}$}           \\
    &           &      &       &       &       &       &        &        \\
    &           & 0.1  & 0.3   &  0.5  & 0.7   & 0.8   & 0.9    & 0.99   \\             
\hline
    &           &      &       &       &       &       &        &        \\
1   &     0     &  26  &   80  &  137  & 203   & 242   & 292    & 369    \\
1   & $10^{-5}$ &  26  &   80  &  137  & 202   & 242   & 291    & 369    \\
1   & 0.002     &  25  &   77  &  133  & 197   & 237   & 285    & 362    \\ 
1   & 0.020     &  26  &   78  &  134  & 199   & 237   & 286    & 362    \\
    &           &      &       &       &       &       &        &        \\
3   &     0     &  40  &  121  &  210  & 312   & 367   & 438    & 548    \\
    &           &  39  &  110  &  179  & 258   & 299   & 342    & 397    \\ 
3   & $10^{-5}$ &  39  &  119  &  204  & 301   & 359   & 429    & 541    \\
    &           &  38  &  107  &  174  & 249   & 293   & 340    & 391    \\ 
3   & 0.002     &  34  &  103  &  178  & 262   & 312   & 375    & 473    \\
    &           &  32  &   92  &  151  & 218   & 255   & 300    & 348    \\
3   & 0.020     &  28  &   89  &  152  & 224   & 267   & 321    & 405    \\
    &           &  29  &   83  &  137  & 199   & 231   & 269    & 312    \\
    &           &      &       &       &       &       &        &        \\
9   &     0     &  62  &  188  &  322  & 473   & 564   & 678    & 860    \\
    &           &  62  &  171  &  279  & 393   & 462   & 532    & 607    \\ 
9   & $10^{-5}$ &  50  &  152  &  260  & 383   & 457   & 548    & 699    \\
    &           &  48  &  132  &  220  & 322   & 378   & 444    & 511    \\ 
9   & 0.002     &  44  &  128  &  220  & 325   & 387   & 465    & 589    \\
    &           &  40  &  111  &  186  & 269   & 316   & 371    & 421    \\ 
9   & 0.020     &  37  &  113  &  195  & 287   & 343   & 411    & 541    \\
    &           &  36  &  102  &  169  & 247   & 291   & 340    & 384    \\
    &           &      &       &       &       &       &        &        \\
20  &     0     &  85  &  256  &  434  & 644   & 766   & 918    & 1147   \\
    &           &  77  &  214  &  349  & 512   & 584   & 672    &  769   \\
20  & $10^{-5}$ &  58  &  177  &  303  & 447   & 532   & 639    &  819   \\
    &           &  55  &  156  &  263  & 386   & 454   & 535    &  626   \\
20  & 0.002     &  49  &  150  &  258  & 380   & 454   & 545    &  686   \\
    &           &  45  &  129  &  221  & 323   & 379   & 444    &  515   \\
20  & 0.020     &  45  &  136  &  233  & 344   & 409   & 490    &  639   \\
    &           &  41  &  118  &  200  & 291   & 342   & 397    &  459   \\
    &           &      &       &       &       &       &        &        \\
60  &     0     & 106  &  316  &  540  & 795   & 947   &1129    & 1423   \\
    &           &  97  &  276  &  476  & 688   & 827   & 952    & 1140   \\
60  & $10^{-5}$ &  75  &  225  &  385  & 566   & 674   & 807    &1017    \\
    &           &  71  &  213  &  364  & 533   & 632   & 751    & 880    \\
60  & 0.002     &  63  &  191  &  326  & 480   & 577   & 687    & 879    \\
    &           &  57  &  175  &  300  & 440   & 517   & 609    & 708    \\
60  & 0.020     &  56  &  171  &  292  & 428   & 509   & 603    & 732    \\
    &           &  50  &  157  &  268  & 389   & 458   & 534    & 600    \\
    &           &      &       &       &       &       &        &        \\            
\hline
\end{tabular}
\end{center}
\end{table} 

\subsection{Velocities on the ZAMS: \boldmath $\upsilon_\mathrm{crit}$, $\upsilon$}

Table~\ref{rotcar} gives for the various models the angular velocity, the period, the polar radius and the equatorial velocity at the critical limit on the ZAMS. Let us note that for obtaining the value of $\Omega_\mathrm{crit}$ or of whatever quantities at the critical limit, it is necessary to compute models at the critical limit. Here we have computed models very near but not exactly at the critical limit, since at that point numerical singularities are encountered. As explained in Sect.~\ref{roche}, using Eq.~(\ref{eq7}) we could have estimated $\Omega_\mathrm{crit}$ from whatever model computed with a lower initial velocity. Doing this for the whole range of initial masses, metallicities and velocities, we would have obtained values within 10\% of those shown in Table~\ref{rotcar}. For the models between 3 and 20 $M_{\sun}$, the error would be always inferior to 3.5\% because in this mass range, $R_\mathrm{p}(\omega)/R_\mathrm{p}(0)$ remains very near 1 (see Fig.~\ref{rprpo}) and Eq. (\ref{eq7}) is a good approximation of Eq. (\ref{eq6}).
For the 1 and 60 $M_{\sun}$ the errors would be larger, amounting to a maximum of 9.3 and 6.7\% respectively, showing the need of using rather Eq.~(\ref{eq6}) if $\omega << 1$. In general the error decreases drastically when $\omega$ increases, so in order to obtain the values given in Table~\ref{rotcar}, we have used our faster rotating model, deriving the value of $\Omega_\mathrm{crit}$ (using Eq.~(\ref{eq7})) with an error of less than 0.001\%. The critical period, $P_\mathrm{crit}$ is then simply given by $P_\mathrm{crit}=2\pi/\Omega_\mathrm{crit}$. The polar radius is obtained from Eq.~(\ref{eq4}) and the critical equatorial velocity is deduced from $\upsilon_\mathrm{crit}=1.5\, R_\mathrm{p,crit}\,\Omega_\mathrm{crit}$.
  \begin{figure*}
  \resizebox{\hsize}{!}{\includegraphics[angle=-90]{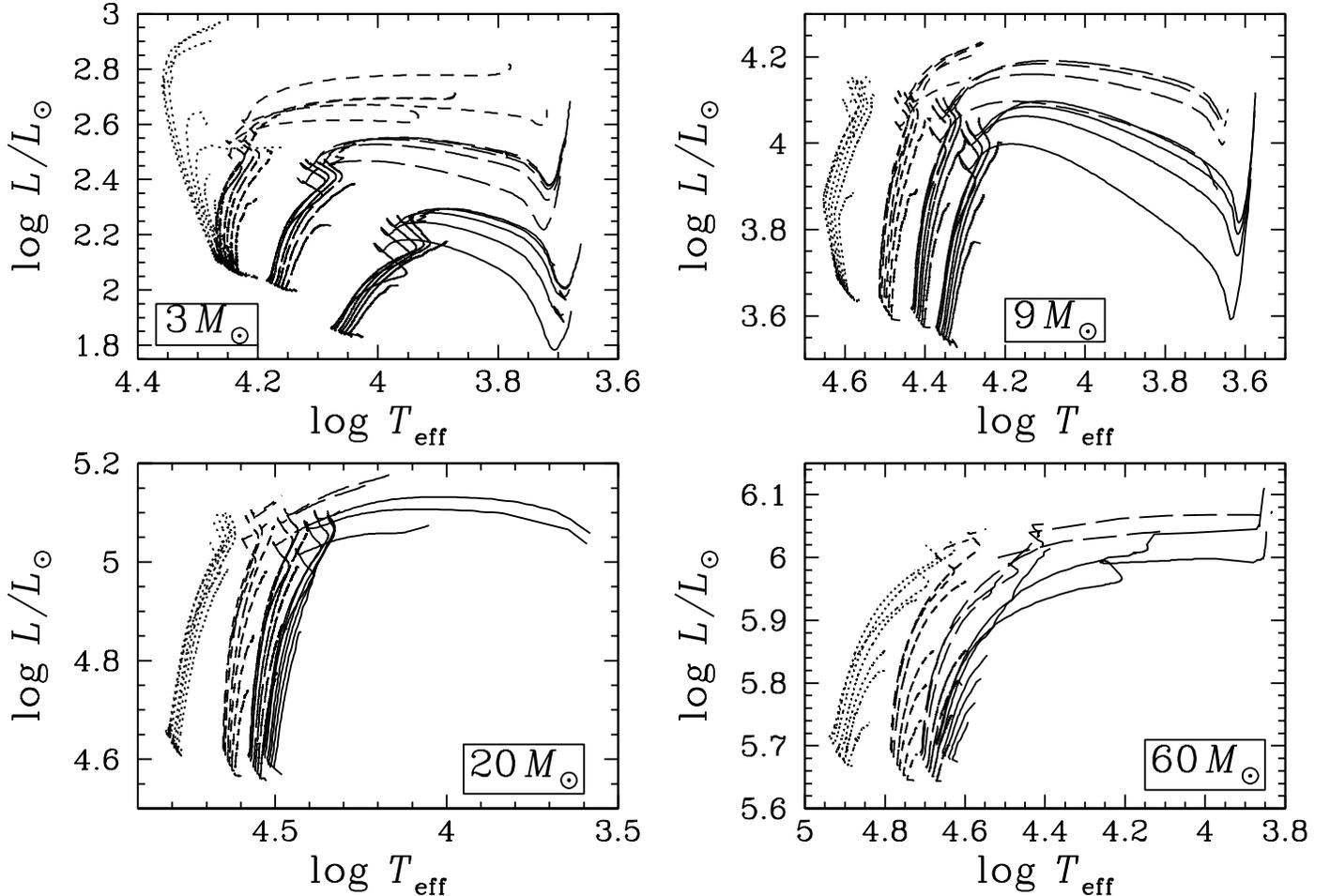}}
      \caption{Evolutionary tracks in the theoretical Hertzsprung-Russell
      diagrams for all the models computed beyond the ZAMS. The effective temperature corresponds to a surface averaged effective temperature. The computations are stopped when
      the star reaches the critical limit or before the beginning of the core
      He-burning phase.}
       \label{dhrrot}
  \end{figure*}

The variation of the critical velocity as a function of the initial mass and metallicity is shown in Fig.~\ref{vcrit}. For a given initial mass, the critical velocity is higher at lower $Z$: \emph{e.g.} for the 60 $M_{\sun}$ model, the critical velocity goes from slightly more than 800 km s$^{-1}$ at $Z=0.020$ to more than 1500 km s$^{-1}$ at $Z=0$. It is a consequence of the smaller radii of stars at low metallicity (see Eq.~(\ref{eq4b})) as can be seen from  Fig.~\ref{rayons}. 

At a given metallicity, the critical velocity becomes higher for higher initial masses. This reflects the fact that the ratio of the mass to the radius increases with the mass of star (cf. Sect. \ref{rlt}). The critical rotation periods are between 2.4 and 25 hours for the whole domain of masses and metallicities explored in the present work. Let us just mention here that, as will be described in Sect.~\ref{bul}, the critical velocity evolves with time, so the values given here are values at the beginning of the star's evolution and should not be used throughout the whole MS phase.

Fig.~\ref{vsurf} shows relations between  $\upsilon$ and $\omega$ on the ZAMS for different initial mass models at various metallicities. The corresponding values are also given in Table~\ref{veloc}. We see that, for a given value of  $\omega$, higher surface velocities are obtained at lower metallicities and for higher initial masses.

%
\section{Evolution on the Main Sequence \label{evolms}}

\subsection{Hertzsprung-Russell diagram and lifetimes}

\begin{figure}
  \resizebox{\hsize}{!}{\includegraphics[angle=-90]{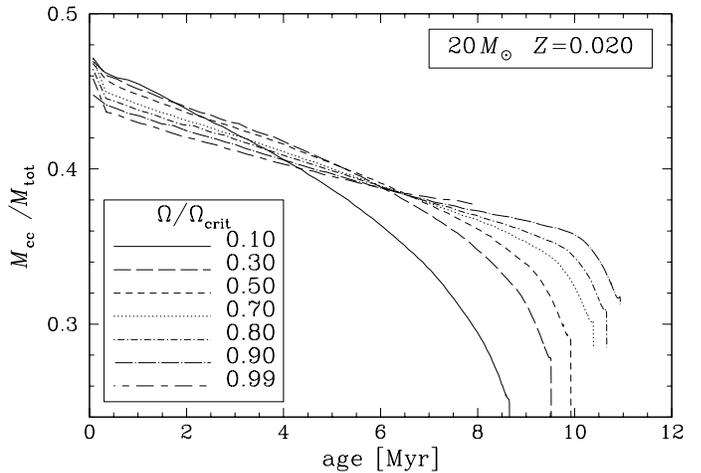}}
      \caption{Evolution of the size of the convective core during MS, for the
      various values of rotational rate, in the 20 $M_{\sun}$ at standard
      metallicity.}
       \label{qcc}
  \end{figure}

Evolutionary tracks in the Hertzsprung-Russell diagram for the present models are shown in Fig.~\ref{dhrrot}. The models have been evolved until either the reaching of $\Omega_\mathrm{crit}$ or just before the onset of central Helium burning. For each mass, the effect of reducing the metal content shifts the tracks towards the blue part of the diagram and, for 3 and 9 $M_{\sun}$, towards higher luminosities. In each metallicity subgroup, the effect of raising up the rotational rate shifts the tracks towards lower luminosities and cooler temperatures on the ZAMS. When the star evolves, rotational mixing continuously refuels the convective core in fresh hydrogen and brings newly produced helium in the radiative zone. The enrichment of the radiative envelope in helium makes the star brighter and bluer by decreasing the opacity. If we look at what happens to the convective core (Fig.~\ref{qcc}), we see that at first, a higher rotational rate leads to a smaller convective core, due to the lower central temperature. Then, we see that the decrease of the core is slower because of the refuelling in hydrogen. At the end of the MS, the core is much larger in the fast rotating models. In that respect rotation acts as a core overshoot.

In Table~\ref{life} the lifetimes of our stellar models are given as well as the time at which the critical limit (if reached) is first encountered during the MS phase. For the models which were not pursued until the end of the MS phase, we give an estimate of the MS lifetime (this is indicated by an asterisk) obtained by inspections of the variation of the central hydrogen mass fraction with time. Note that we suppose that the star, once it reaches the critical limit, remains in the vicinity of this limit. This behaviour has been confirmed by previous computations in which models were pursued beyond the point where they first reach the critical limit \citep[see][]{EkTar}. Most of the 60 $M_{\sun}$ stellar models encounter the critical limit very early. In that case only the time of reaching the limit is given. We see that for $\omega=0.7$ the MS lifetime is increased by 15 - 23\% (depending on the initial mass and metallicity) in comparison with the models with $\omega=0.1$.
\begin{table}[t]
      \caption{Lifetimes (in Myr) during the MS phase. An asterisk means that
      the value was estimated (see text).
} \label{life}
\begin{center}\scriptsize
\begin{tabular}{cccccccc}
\hline \hline
          &        &         &         &        &        &         &        \\
 $Z$      & \multispan7 $\Omega/\Omega_\mathrm{c}$           \\
          &        &         &         &        &        &         &        \\
          & 0.1    & 0.3     &  0.5    & 0.7    & 0.8    & 0.9     & 0.99   \\
\hline
          &        &         &         &        &        &         &         \\
\multispan8 {3$M_{\sun}$} \\     
          &        &         &         &        &        &         &         \\
    0     & 215.6  &  227.8  &  237.8  & 249*   & 255*   & 264*    & 275*    \\
          &        &         &         & (223.7)& (206.2)& (189.0) & (170.6) \\
$10^{-5}$ & 224.9  &  243.6  &         & 261.5  & 265.6  & 271*    & 271*    \\
          &        &         &         &        &        & (270.6) & (213.4) \\
0.002     & 274.2  &  301.3  &  312.3  & 320.9  & 325.4  & 331*    & 331*    \\
          &        &         &         &        &        & (316.4) & (237.9) \\
0.020     & 357.0  &  399.6  &  424.4  & 439.2  & 444.4  & 454*    & 484*    \\
          &        &         &         &        &        & (446.7) & (302.9) \\
          &        &         &         &        &        &         &         \\
\multispan8 {9$M_{\sun}$} \\            
          &        &         &         &        &        &         &         \\
    0     &  20.31 &  22.11  &  23.24  & 24.31  & 24.74  & 25*     & 25*     \\
          &        &         &         &        &        & (24.77) & (16.12) \\
$10^{-5}$ &  26.62 &  28.70  &  29.98  & 30.85  & 31.29  & 31.8*   & 32.4*   \\
          &        &         &         &        &        & (31.30) & (25.03) \\
0.002     &  27.77 &  30.11  &  31.32  & 32.26  & 32.87  & 33.5*   & 34.0*   \\
          &        &         &         &        &        & (31.84) & (26.00) \\
0.020     &  27.59 &  30.33  &  31.62  & 32.62  & 33.0*  & 34.0*   & 34.0*   \\
          &        &         &         &        & (32.91)& (28.89) & (23.05) \\
          &        &         &         &        &        &         &         \\
\multispan8 20$M_{\sun}$ \\            
          &        &         &         &        &        &         &         \\
    0     &  8.003 &  8.554  & 8.868   & 9.199  & 9.324  & 9.434   & 9.44*   \\
          &        &         &         &        &        &         & (8.616) \\
$10^{-5}$ &  9.099 &  9.853  & 10.213  & 10.54* & 10.68* & 10.73*  & 10.73*  \\
          &        &         &         &(10.533)& (9.626)& (7.843) & (5.651) \\
0.002     &  9.303 & 10.117  & 10.527  & 10.85* & 11.00* & 11.06*  & 11.15*  \\
          &        &         &         &(10.840)&(10.033)& (8.444) & (6.712) \\
0.020     &  8.663 &  9.521  &  9.928  & 10.39* & 10.66* & 10.96*  & 10.96*  \\
          &        &         &         &(10.389)&(10.662)&(10.937) & (7.892) \\
          &        &         &         &        &        &         &         \\
\multispan8 60$M_{\sun}$ \\            
          &        &         &         &        &        &         &         \\
   0      &  3.633 &  3.802  &  3.924  &        &        &         &         \\
          &        &         &         & (3.386)& (2.538)& (2.259) & (1.173) \\
$10^{-5}$ &  3.874 &  4.112  &         &        &        &         &         \\
          &        &         & (3.696) & (2.746)& (2.200)& (1.525) & (0.783) \\
0.002     &  3.994 &  4.251  &         &        &        &         &         \\
          &        &         & (4.432) & (2.818)& (2.336)& (1.717) & (1.087) \\
0.020     &  3.714 &  3.978  &         &        &        &         &         \\
          &        &         & (3.960) & (2.426)& (2.031)& (1.595) & (1.179) \\
          &        &         &         &        &        &         &         \\
\hline
\end{tabular}
\end{center}
\end{table}

\subsection{Evolution of the equatorial velocity \label{evolveq}}

The evolution of the rotation velocities at the stellar surface depends of course directly on the movements of expansion/contraction of the star. But besides this, it also strongly depends on two factors: the internal mechanism of angular momentum transport (internal coupling) and the mass loss. 

An extreme case of internal angular momentum transport is the one which imposes solid body rotation at each time in the course of the evolution of the star. A strong coupling is then realized between the contracting core and the expanding envelope. The spinning up of the central regions is transmitted to the stellar surface. Thus, during the MS phase, the surface velocity approaches the critical limit and may even reach it, depending on the initial conditions \citep[see][]{Sa70,La97,Sap07}. Another extreme case is the case of no transport of angular momentum. Each stellar layer keeps its own angular momentum. The variation of $\Omega$ is then simply governed by the local conservation of the angular momentum and, at the surface, $\Omega \sim R^{-2}$. When the radius increases, the surface angular velocity decreases more rapidly than the classical critical angular velocity (see Eq.~(\ref{eq4})), thus the star evolves away from the critical limit.
In the present models, without magnetic fields, the situation is intermediate between these two extreme cases. A moderate coupling is exerted mainly by meridional circulation, which is more efficient than shear transport, as far as transport of angular momentum is concerned \citep[see \emph{e.g.}][]{MMV}. 

The second important effect governing the evolution of angular momentum is mass loss. The strong mass losses during the MS phase, expected for high mass stars at high metallicities, decrease the angular momentum of the stars. The importance of the decrease of the angular momentum depends on how matter is ejected. In the extreme case where all the matter would be ejected along narrow jets centred on the rotational axis, nearly no angular momentum would be lost, while if all the matter is ejected near the equator plane, this maximizes the amount of angular momentum lost. Depending on the geometry of the wind, the surface velocity may thus decrease or increase. In the present work, we did not account for the effects of wind anisotropies (which would be important only for a relatively narrow range of masses and velocities, see \citealt{MMX}), thus, in the present case, strong mass loss always leads to a decrease of the surface velocity. 
\begin{figure}
  \resizebox{\hsize}{!}{\includegraphics{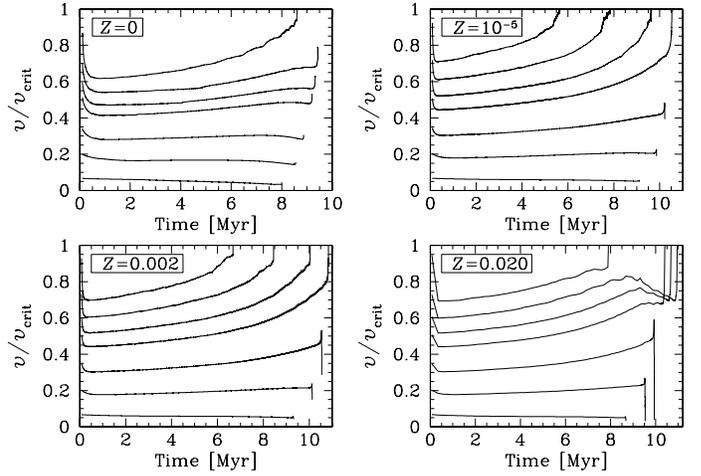}}
      \caption{Evolution of the ratio $\upsilon/\upsilon_\mathrm{crit}$ during the
      core hydrogen-burning phase for the 20 $M_{\sun}$ models at various
      metallicities. The computations were stopped as soon as the star reaches
      the critical limit or just before the beginning of the core He-burning phase.}
       \label{vevol}
  \end{figure}

Figure~\ref{vevol} shows the evolution of the surface velocity expressed as a fraction of the critical velocity during the MS phase for 20 $M_{\sun}$ stellar models. Quite generally the curves can be decomposed into three parts, whose importance depends on the initial rotation and metallicity:
\begin{enumerate}
\item At the very beginning, there is a short adjustment period, which lasts for a few percent of the MS lifetime, during which meridional circulation transports angular momentum from the outer parts of the star to the inner ones. This slows down the surface of the star. Then, in the interior, shear turbulence becomes active and erodes the gradients built by the meridional circulation. Under the influence of these two counteracting effects the $\Omega$-profile converges toward an equilibrium configuration
\citep{Za92,DIW,MMV}. From Fig.~\ref{vevol}, one sees that for lower initial values of $\upsilon/\upsilon_\mathrm{crit}$ the decrease is much weaker than for higher values. It may even completely disappear. It illustrates the dependence of the meridional velocity on $\Omega$. 
This feature also shows that, in the present theoretical context, we cannot have stars which would remain near the critical limit starting from the ZAMS. Indeed, our most extreme case, the model with $\Omega/\Omega_\mathrm{crit}=0.99$  evolves away from the critical limit just after the ZAMS. Of course we can wonder to what extent this behaviour is dependent on the chosen initial conditions. For instance would this behaviour be the same, if one would have computed realistic pre-MS evolution? Probably in that case the model would not have solid body rotation on the ZAMS and thus the above feature is in part artificial. It is however difficult to say more due to the complexity of the processes affecting the angular momentum of stars during their formation process. The point which has to be kept in mind is that 
{\it in the present models we cannot have non-evolved critically rotating stars.} Let us stress that these stars should be anyway very rare and their absence should not have any impact on the frequency of stars predicted to rotate at or near the critical velocity. The above effect implies also that the velocities on the ZAMS given in Table~2 are probably quantities not directly comparable to observed ones. Indeed, in the most optimistic case, where ZAMS models would indeed rotate as solid body, the chance to see stars at the very beginning of their relaxation period is very small. More realistic values of the initial velocities would correspond to time averaged values computed over the period during which about ten percent of the initial content in hydrogen has been burnt in the core. These values are also given in Table~\ref{veloc}.
It is important to note that models treating advection as a diffusion would miss this first relaxation period.
Indeed during this phase, meridional currents build the gradient of $\Omega$, a process that diffusion cannot do, diffusion always eroding any gradient.
\item As explained for instance in \citet{MMAA}, after the first adjustment period, a large outer cell of meridional circulation sets in, transporting angular momentum from the inner parts of the star to the outer ones\footnote{In our model, during the MS phase, we have two cells of meridional circulation, an inner one where matter moves clockwise, and an outer one where matter moves counter-clockwise. The velocities in the outer cell are much larger than the velocities of the inner cell. As a consequence, the outer cell clearly dominates the transport of the angular momentum in the star, transporting it from inner regions towards outer ones.}. If this transport is rapid enough and the stellar winds not too intense, then $\upsilon/\upsilon_\mathrm{crit}$ increases (an important distinction has to be made here: it is the ratio $\upsilon/\upsilon_\mathrm{crit}$ which increases, not necessarily the equatorial velocity itself, see  Fig.~\ref{vvvevol} and Sect. \ref{bul} for discussion on this point). It is the case for all the models with $\omega \ge 0.3$ at $Z \ge 0.00001$. For metal-free stars, the increase of $\upsilon/\upsilon_\mathrm{crit}$ only occurs for the models with $\omega \geq 0.7$. This is a consequence of the Gratton-\"Opik effect (see the term in brackets in Eq.~(\ref{gratton}) making the meridional circulation velocity smaller at lower $Z$.
\item At the end of the MS phase, the increase of $\upsilon/\upsilon_\mathrm{crit}$ accelerates when the star contracts after the blue hook in the HR diagram. The rapid change of the velocity at that stage is due to the rapid change of the radius, not to any transport mechanisms.
\end{enumerate}

Looking at the right bottom panel of Fig.~\ref{vevol} (models at $Z=0.020$), another interesting effect can be discussed. We see for instance that the 20 $M_{\sun}$ stellar models with initial value of $\upsilon/\upsilon_\mathrm{crit}$ between 0.5 and 0.7 ($\omega$ between 0.7 and 0.9) show a bump around an age of 9 Myr. This occurs because at that time, for these models, mass loss increases enough to remove efficiently the angular momentum brought to the surface by the meridional currents. The surface velocity thus evolves away from the critical limit, reaching it again only at the very end of the MS phase. This feature is not seen at lower metallicity because mass loss rates are lower. This trend was already put in evidence by \citet{MMVII}. For a given degree of coupling, {\emph{the mass loss rates play a most critical role in the evolution of the surface rotation}}.
\begin{figure}
  \resizebox{\hsize}{!}{\includegraphics[angle=-90]{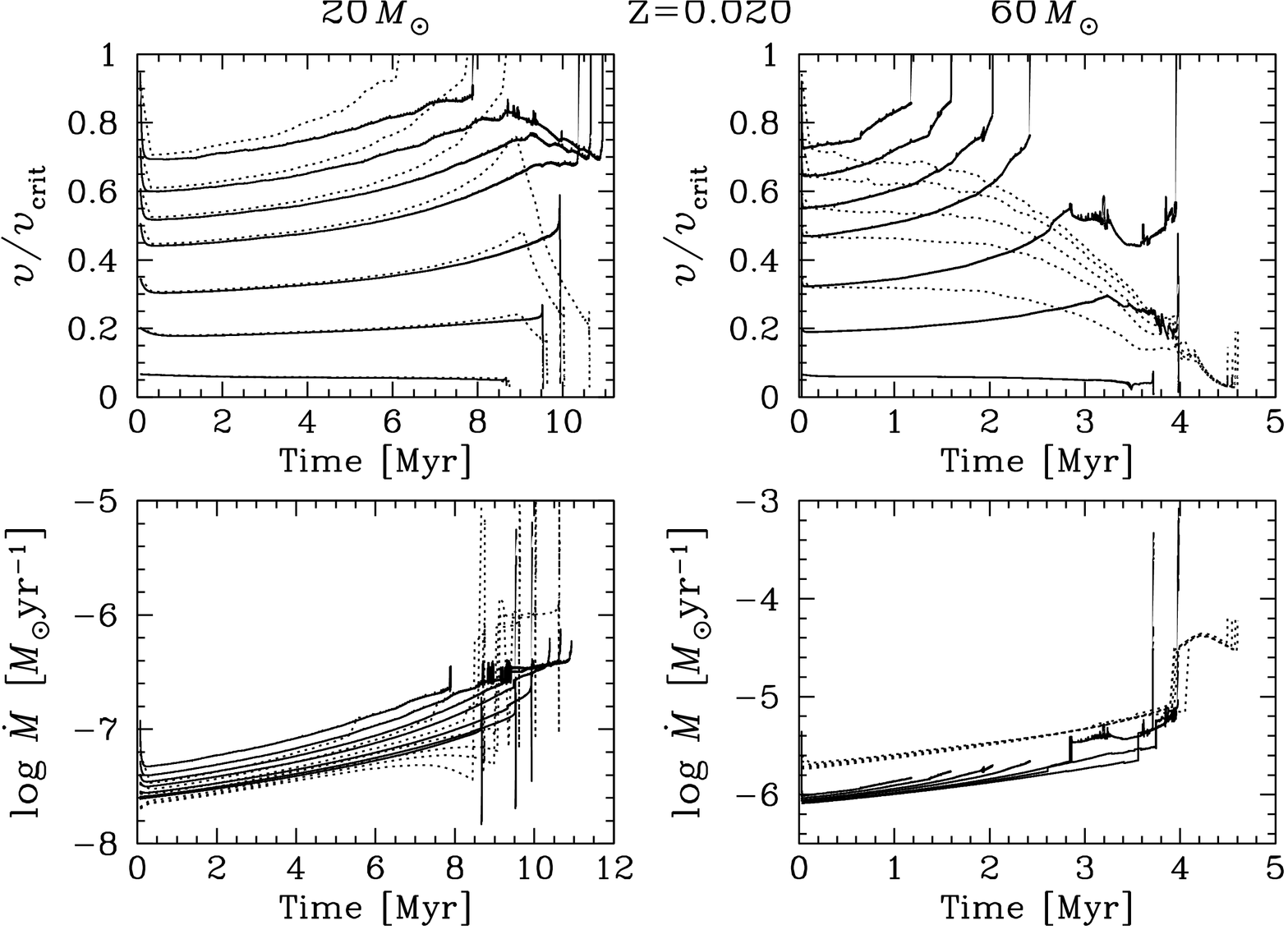}}
      \caption{Comparison of the evolution of $\upsilon/\upsilon_\mathrm{crit}$ (top) and of $\log \dot{M}$ (bottom) for the 20 $M_{\sun}$ models (left) and the 60 $M_{\sun}$ models (right) at standard metallicity, when using \citealt{kupu00} (continuous lines) or \citealt{Vink00} (dotted lines).}
       \label{kudrvinkcomp}
  \end{figure}

In order to check the validity of our results, we have computed another set of 20 and 60 $M_{\sun}$ models at $Z=0.020$ with the mass loss prescription of \citealt{Vink00} \citep[hereafter][]{Vink00}. In the case of the 20 $M_{\sun}$ models, the mass loss rates obtained in the new set are slightly lower than the ones obtained with \citealt{kupu00} \citepalias{kupu00}. The $\omega \geq 0.8$ models meet thus the break-up limit earlier in their evolution. The $\omega=0.7$ model shows also a bump in the velocity, as described above, but then it stays far from break-up limit, even at the end of the MS unlike what was the case with the \citetalias{kupu00} rates. This has for consequence that the minimum $\omega$ required to reach the break-up limit during evolution will be slightly higher with \citetalias{Vink00} rates, while the fast rotators will spend a little more time at break-up limit (Fig.~\ref{kudrvinkcomp}, left panels). But the qualitative result is generally similar to the one obtained with \citetalias{kupu00} rates. In the 60 $M_{\sun}$ case, the opposite happens: \citetalias{Vink00}'s rates are much stronger (by a factor 2 during the MS), and thus no model reaches the break-up limit anymore, even the one starting its evolution with $\omega=0.99$ (Fig.~\ref{kudrvinkcomp}, right panels). The most massive stars' evolution is dramatically influenced by the choice of the mass loss rates and it can be concluded that in that case, the comparison between theoretical models and observations constitutes rather a check for the mass loss rates than anything else. 

Coming back to our initial set of models, we present in Table~\ref{tvmoy} the time averaged velocities during the MS phase or during the period before the reaching of the critical limit. This table is useful in order to connect the values of $\omega$ on the ZAMS with observable rotational properties during the MS phase (see Sect.~\ref{bul}). 
\begin{table}[t]
\caption{Time averaged equatorial velocities in km s$^{-1}$ corresponding to
various ZAMS values of $\omega=\Omega/\Omega_\mathrm{crit}$. The average is
taken either on the whole MS phase or on the part of the evolution before the
critical limit is reached. For these latter models, the mass fraction of
hydrogen at the centre when the surface velocity reaches the critical limit is
indicated in parenthesis.} \label{tvmoy}
\scriptsize
\begin{tabular}{ccccccccc}
\hline \hline
M   &  $Z$      & & &  $\omega$ & = &$\Omega/\Omega_\mathrm{crit}$           \\
    &           &      &       &       &       &       &        &        \\
    &           & 0.1  & 0.3   &  0.5  & 0.7   & 0.8   & 0.9    & 0.99   \\
\hline
    &           &      &       &       &       &       &        &        \\
3   &     0     &  38  &  108  &  198  & 294   & 333   & 373    & 423    \\
    &           &      &       &       & (5e-2)& (0.1) & (0.17) & (0.26) \\
3   & $10^{-5}$ &  33  &   95  &  167  & 247   & 297   & 353    & 416    \\
    &           &      &       &       &       &       & (7e-3) & (0.30) \\
3   & 0.002     &  27  &   80  &  139  & 207   & 246   & 298    & 352    \\
    &           &      &       &       &       &       & (0.1)  & (0.38) \\
3   & 0.020     &  23  &   69  &  120  & 181   & 215   & 258    & 311    \\
    &           &      &       &       &       &       & (3e-2) & (0.38) \\
    &           &      &       &       &       &       &        &        \\
    &           &      &       &       &       &       &        &        \\
9   &     0     &  56  &  151  &  265  & 394   & 477   & 569    & 666    \\
    &           &      &       &       &       &       & (1e-5) & (0.35) \\
9   & $10^{-5}$ &  39  &  117  &  204  & 305   & 363   & 435    & 515    \\
    &           &      &       &       &       &       & (3e-2) & (0.29) \\
9   & 0.002     &  33  &   99  &  173  & 256   & 305   & 369    & 425    \\
    &           &      &       &       &       & (3e-4)& (9e-2) & (0.31)\\
9   & 0.020     &  30  &   90  &  157  & 236   & 284   & 341    & 387    \\
    &           &      &       &       &       & (2e-2)& (0.21) & (0.36) \\
    &           &      &       &       &       &       &        &        \\    
    &           &      &       &       &       &       &        &        \\
20  &     0     &  54  &  169  &  295  & 446   & 507   & 588    & 721    \\
    &           &      &       &       &       &       &        & (0.13) \\
20  & $10^{-5}$ &  45  &  144  &  254  & 389   & 469   & 557    & 651    \\
    &           &      &       &       & (3e-3)& (0.15)& (0.33) & (0.48) \\
20  & 0.002     &  37  &  120  &  213  & 323   & 387   & 456    & 528    \\
    &           &      &       &       & (2e-3)& (0.14)& (0.30) & (0.43)\\
20  & 0.020     &  32  &  107  &  189  & 279   & 325   & 369    & 447    \\
    &           &      &       &       & (2e-4)& (1e-4)& (6e-4) & (0.30) \\
    &           &      &       &       &       &       &        &        \\
    &           &      &       &       &       &       &        &        \\
60  &     0     &  63  &  248  &  456  & 645   & 830   & 930    & 1167    \\
    &           &      &       &       & (0.20)& (0.38)& (0.45) & (0.61) \\
60  & $10^{-5}$ &  63  &  221  &  408  & 587   & 687   & 797    & 890    \\
    &           &      &       & (0.16)& (0.37)& (0.47)& (0.57) & (0.67) \\
60  & 0.002     &  48  &  179  &  323  & 474   & 552   & 641    & 727    \\
    &           &      &       & (5e-5)& (0.40)& (0.45)& (0.54) & (0.63)\\
60  & 0.020     &  37  &  140  &  258  & 400   & 472   & 547    & 609    \\
    &           &      &       & (8e-2)& (0.38)& (0.44)& (0.51) & (0.56) \\
    &           &      &       &       &       &       &        &        \\      
\hline
\end{tabular}
\end{table}

\subsection{Surface enrichment}
 \begin{figure*}
 \includegraphics[height=.925\textheight]{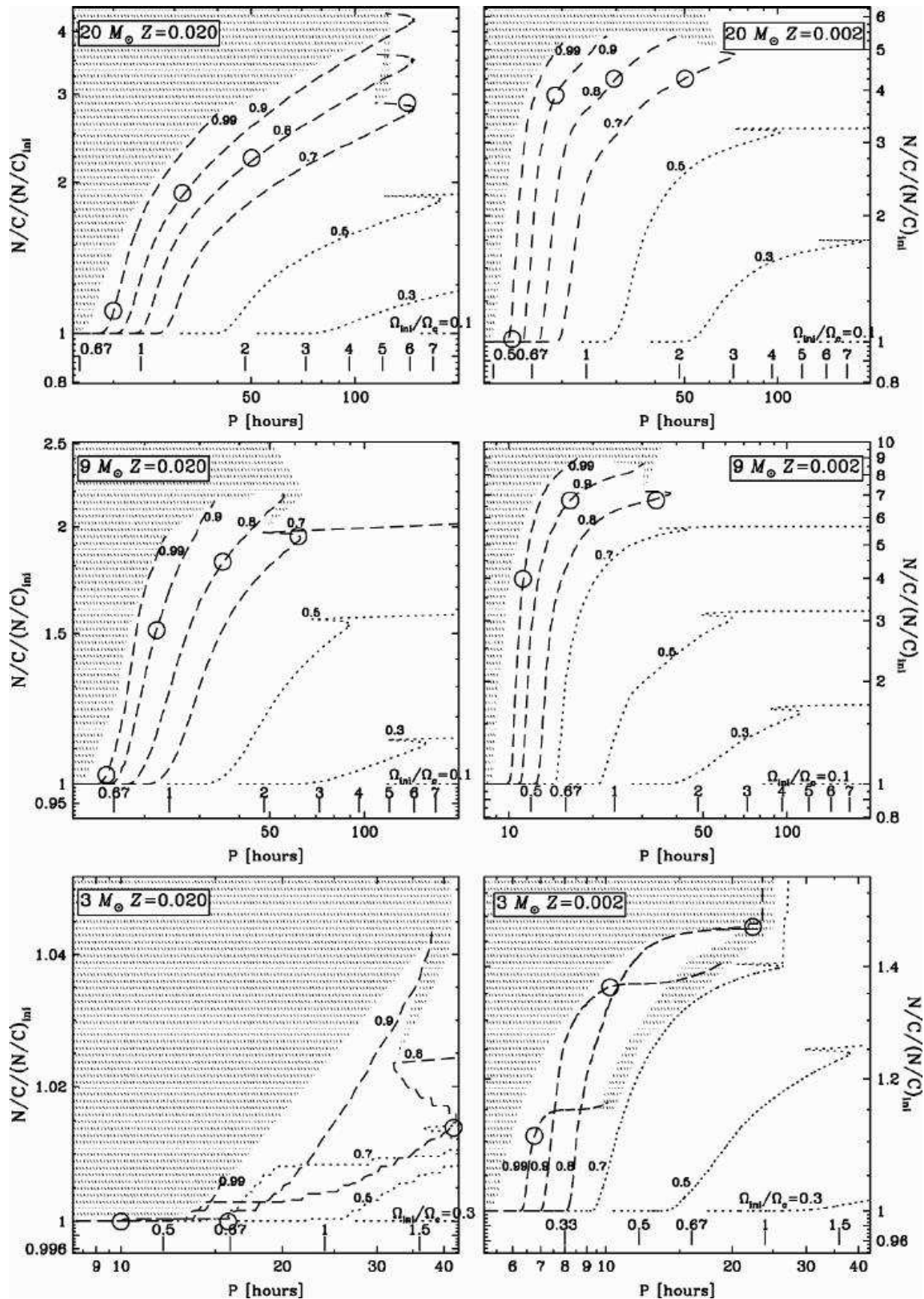}
      \caption{Evolutionary tracks in the plane surface N/C ratio, normalized to its initial value, versus the rotational period in hours for different initial mass stars, various initial velocities and for the metallicities $Z$=0.02 and 0.002. Positions of some periods in days are indicated at the bottom of the figure. The dotted tracks never reach the critical limit during the MS phase. The short dashed tracks reach the critical limit during the MS phase. The dividing line between the shaded and non-shaded areas corresponds to the entrance into the stage when the star is at the critical limit during the MS evolution. Big circles along some tracks indicate the stage when $\upsilon/\upsilon_\mathrm{crit}$ becomes superior to 0.7. If Be stars are rotating at velocities superior to 70\% of the critical velocity, present models would predict that they would lie in the region comprised between the big circles and the dividing line. Beware the different vertical scales used when comparing similar masses at different metallicities.}
       \label{ncp}
  \end{figure*}

In Fig.~\ref{ncp} the tracks are plotted in the plane ${\rm (N/C)/(N/C)_{ini}}$ versus $P$, where $P$ is the rotational period in hours. During the evolution the surface is progressively enriched in CNO burning
products, {\it i.e.} enriched in nitrogen and depleted in carbon. At the same time, the rotational period  increases. The hatched regions corresponds to zones where stars would rotate at or faster than the critical limit. We can note the following features:
\begin{itemize}
\item As already mentioned in previous works \citep[see e.g.][]{MMVII}, for a given value of the initial mass, of $\omega$ and for a given evolutionary stage, the surface enrichments are stronger in metal poor stars. This is related mainly to the fact that the gradients of $\Omega$ are steeper in metal poor stars due the slower meridional currents in more compact stars.
We note that surface enrichments for 3 $M_{\sun}$ stellar models at $Z=0.02$ remain in the range of a few percents, even for high rotation rates.
This means that for these low mass stars, observations will probably not be able to see such small enrichments.
The same initial mass models at $Z=0.002$ present enhancements of up to 45\%. 
\item Also the surface enrichments are higher in more massive stars, a feature which is due to higher efficiency of the shear mixing when the initial mass increases. 
\item Stars with higher initial velocities reach higher N/C ratios. Let us recall here that the computations were stopped when the critical limit is reached, this explains why for instance, the maximum surface enrichment indicated for the 9 $M_{\sun}$ stellar model at $Z=0.02$ is lower for $\Omega/\Omega_\mathrm{crit}=0.99$ than for $\Omega/\Omega_\mathrm{crit}=0.8$ and 0.9. Would the model with $\Omega/\Omega_\mathrm{crit}=0.99$ be continued, it would reach higher values than those shown in Fig.~\ref{ncp}.
\item In general, stars near the critical limit show nitrogen enhancements. There are however cases where no or very small enhancements would be expected. These cases are: the stars starting their evolution with very high rotation rates and still being nearly non-evolved. 
For instance, in case of the 9 M$_\odot$ stellar model at $Z=0.02$ with $\Omega/\Omega_\mathrm{ini} = 0.99$, $\upsilon/\upsilon_\mathrm{crit}$ becomes superior to 0.7 while still no significant surface enrichment in nitrogen has occurred. Another case where no observable surface enrichment is expected is for stars at $Z=0.02$ and with initial masses around 3 M$_\odot$ (see lower left panel). 
\item No stars with masses between 3 and 20 $M_{\sun}$ and with rotation periods longer than about 5 days are expected to be at the critical limit. 
\end{itemize}

\section{Initial conditions for stars reaching near critical velocities \label{bul}}
%
\begin{figure}
  \resizebox{\hsize}{!}{\includegraphics{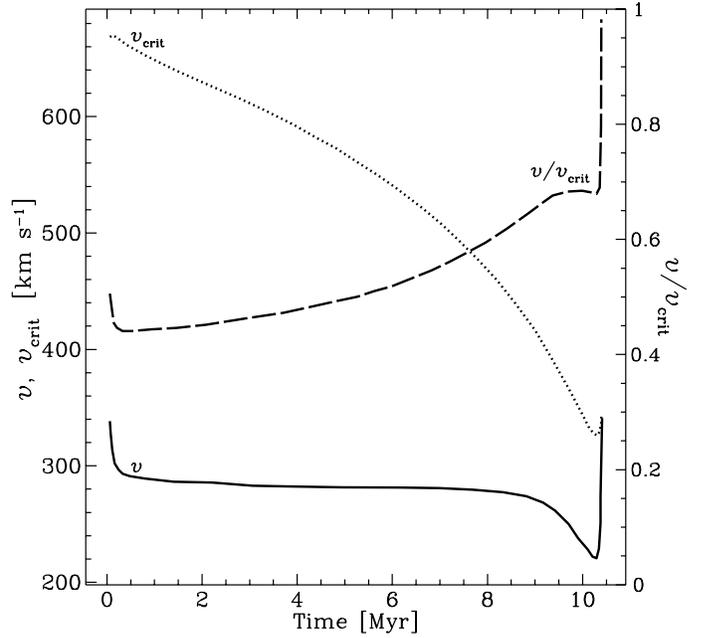}}
      \caption{Evolution of the equatorial velocity $\upsilon$ (continuous line), the critical velocity $\upsilon_\mathrm{crit}$ (dotted line) and their ratio (dashed line) for the 20 $M_{\sun}$ at standard metallicity with $\omega=0.70$ initially ($\upsilon/\upsilon_\mathrm{crit}=0.50$).}
       \label{vvvevol}
  \end{figure}

Let us first recall that an important ingredient of the ratio $\upsilon/\upsilon_\mathrm{crit}$ is the behaviour of $\upsilon_\mathrm{crit}$ itself. Fig.~\ref{vvvevol} shows how it evolves in time in the case of the 20 $M_{\sun}$ model at standard metallicity with $\omega=0.7$ initially ($\upsilon/\upsilon_\mathrm{crit}=0.5$). During evolution, $\upsilon_\mathrm{crit}$ gets lower, and in the present case the final value amounts to only half the initial one. It can be explained by the modifications of the stellar parameters: some mass is lost and the radius is slowly inflating, so the surface gravity gets lower and can be more easily counterbalanced by the centrifugal force. In the example shown in Fig.~\ref{vvvevol}, we see that even though the equatorial velocity is slowly decreasing during the MS, the ratio $\upsilon/\upsilon_\mathrm{crit}$ increases steadily. It is important to take into account this evolutionary effect when one wants to determine the $\upsilon/\upsilon_\mathrm{crit}$ ratio at a given time. Keeping the initial value would lead to a completely wrong result. Let us mention that in that perspective, our results are not contradictory with the observations of \citet{HG06b} or \citet{Abt03} who find a clear spin down with evolution for the stars in their sample. Except the most extreme ones with $\omega=0.99 - 0.90$, all our standard metallicity models in the mass range $3 - 9 M_{\sun}$ show a steady or decreasing equatorial velocity, though their $\upsilon/\upsilon_\mathrm{crit}$ ratios increase.

\begin{figure*}
  \resizebox{\hsize}{!}{\includegraphics[angle=-90]{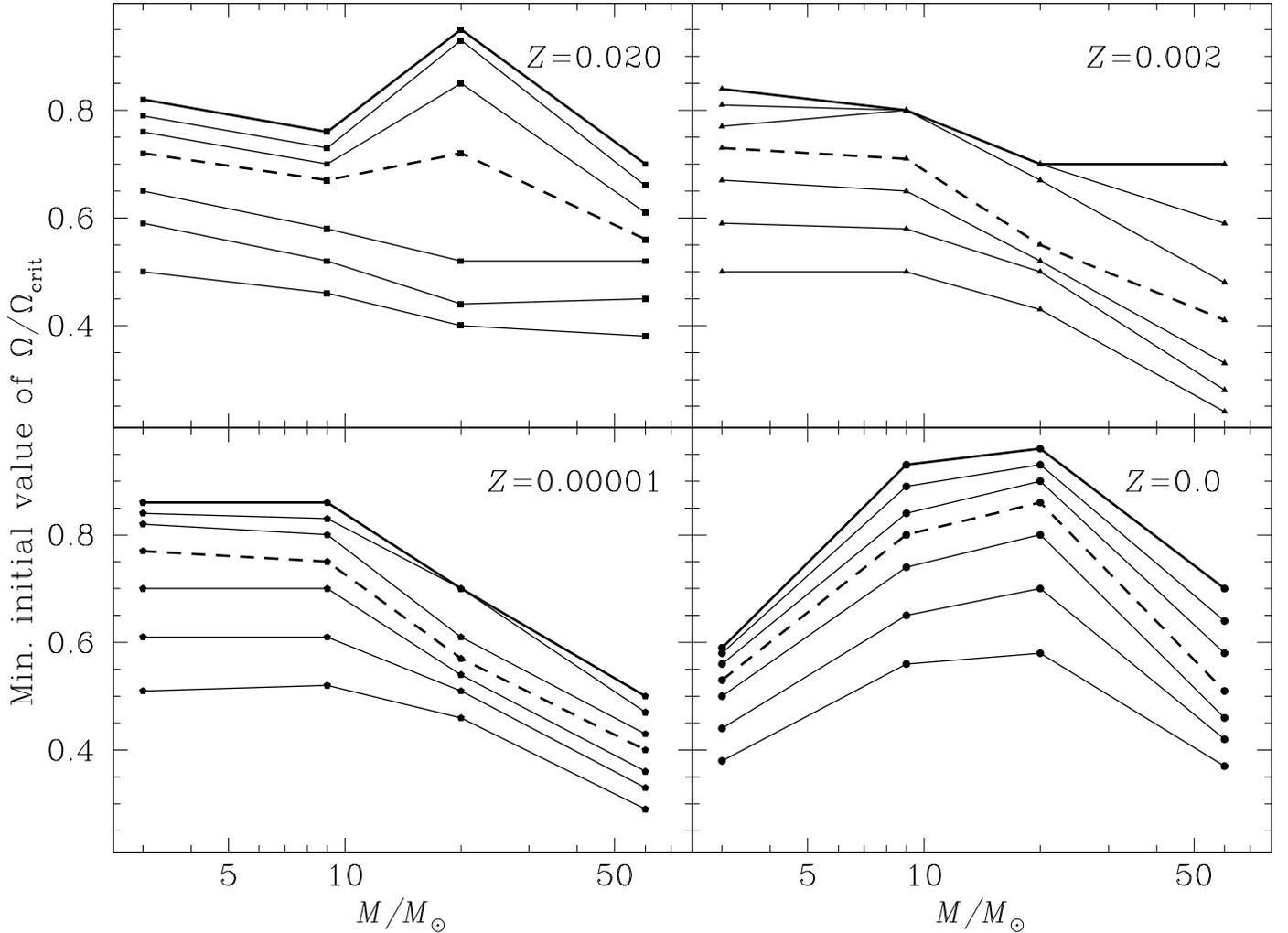}}
      \caption{Minimum values of $\omega=\Omega/\Omega_\mathrm{crit}$ on the ZAMS required for reaching
      during the MS phase values of $\upsilon/\upsilon_\mathrm{crit}$ superior to 0.4 (bottom line), 0.5, 0.6, 0.7 (dashed), 0.8, 0.9 and 0.99 (top heavy line) respectively. The initial metallicity considered is indicated on each panel.}
       \label{fvlimite}
  \end{figure*}

From our computations, we can determine for each model specified by its initial mass and metallicity, the values of $\Omega/\Omega_\mathrm{crit}$ on the ZAMS required to reach given limits of $\upsilon/\upsilon_\mathrm{crit}$ during the MS phase. These values are plotted in Fig.~\ref{fvlimite}. Using Table~\ref{veloc}, it is possible to translate these limits on $\Omega/\Omega_\mathrm{crit}$ into limits on the equatorial velocities on the ZAMS or at the beginning of the MS phase. Using Table~\ref{tvmoy} a similar correspondence can be obtained with values of the averaged velocity during the MS phase or during the phase before the reaching of the critical velocity.
  
At the standard metallicity, we note that in general the values of $\Omega/\Omega_\mathrm{crit}$ needed on the ZAMS to reach given values of $\upsilon/\upsilon_\mathrm{crit}$ decrease when the mass increases. This comes from the fact that meridional currents are more rapid in the less dense envelope of more massive stars. 
Let us recall that in the outer layer, the dominant term in the expression for the radial component of the meridional circulation $U(r)$ is mainly (Gratton-\"Opik term)
\begin{equation}
\label{gratton}
U(r)\sim  \left[ 1-\frac{\Omega^2}{2\pi G \bar{\rho}} \right] \left(
\frac{\Omega^2 r^3}{GM} \right)
\end{equation}
where $\bar{\rho}$ means the average over the considered equipotential. Thus the outer cell of the meridional circulation will be more active in less dense envelopes.
In Fig.~\ref{fvlimite} upper left panel, the bump around the 20 $M_{\sun}$ stellar model comes from the fact that this model suffers significant mass loss at the end of its MS evolution. This disfavours the reaching of near critical values (except at the very end of the MS evolution, see Fig.~\ref{vevol}). In the case of the 60 $M_{\sun}$ model, when the mass loss rates of \citetalias{kupu00} are used, the near critical values occur well before the star undergo similar enhancements of the mass loss rate and thus, the minimum values of $\Omega/\Omega_\mathrm{crit}$ are lower than for the 20 $M_{\sun}$. The situation is quite different if the mass loss rates from \citetalias{Vink00} are used (see Fig.~\ref{kudrvinkcomp}).

For metallicities $Z=0.002$ and 0.00001, we find that the minimum values of $\Omega/\Omega_\mathrm{crit}$ required to reach near critical velocities also decrease when the initial mass increases. 
More massive stars thus reach more easily the critical limit. 
We note also that for the 60 $M_{\sun}$ at $Z=0.002$, the minimum value of $\Omega/\Omega_\mathrm{crit}$ to reach the critical velocity is around 0.7, while at $Z=0.00001$ it is lowered to values equal to 0.5. 
This is mainly due to the weak radiative winds of the metal-poor models. Since radiative winds are triggered by metal lines, a low metal content implies low mass loss rates. We have already seen in Sect.~\ref{evolveq} that there is a competition between the efficiency of the mass loss (removing some angular momentum and thus decelerating the surface) and the efficiency of the meridional circulation (bringing the angular momentum from the core to the surface and thus accelerating the surface). From our results, we can say that even though the efficiency of the meridional circulation is lower at low $Z$, the effect of lowering the mass loss rates plays the major role and the total budget is in favour of a significant acceleration of the surface, at least down to $Z=0.00001$. Let us note that at some points, the curves are superposed: for instance for the 9 $M_{\sun}$ at $Z=0.002$, the curve for reaching the critical limit, and those for reaching values of $\upsilon/\upsilon_\mathrm{crit}\ge 0.8$ and 0.9 go through the same point. This simply reflects the fact that when such a star reaches the values $\upsilon/\upsilon_\mathrm{crit}\ge 0.8$, it then reaches also the critical limit.

For metal-free stars, one notes that for $M > 9\ M_\odot$, the minimum values of $\Omega/\Omega_\mathrm{crit}$ to approach critical velocities during the MS evolution are in general higher than at higher metallicities. This reflects the greater compactness of these stars. The metal-free 3 $M_{\sun}$ models on the contrary reach much more easily the critical limit than the models at higher metallicities. We can see that their tracks in the HR diagram (see Fig.~\ref{dhrrot}) evolve quite blueward. 
This evolution is due to the fact that metal-free stars in this mass domain must contract over a great part of their MS evolution in order to compensate for the lack of CNO elements. In such a situation, the evolution of $\Omega$ is mainly governed by the local conservation of the angular momentum (nearly no coupling) which favours the approach of the critical limit when the star contracts.
Typically for a 3 $M_{\sun}$ stellar model, the minimum value of $\Omega/\Omega_\mathrm{crit}$ is around 0.6 at $Z=0$ for reaching the critical limit during the MS phase, while it is higher than 0.8 at standard metallicity.

\section{Rotation of stars in clusters of various ages and metallicities \label{freqbe}}

Let us call $n(t,w)$ the number of MS stars in a cluster of age $t$ having surface rotational velocities such that $\upsilon/\upsilon_\mathrm{crit} \ge w$. We only consider stars brighter than about two magnitudes below the turn off and call $N(t)$ the total number of MS stars in the same cluster, in the same interval of magnitudes. The ratio $n(t,w)/N(t)$ can be expressed in the following way
$${n(t,w)\over N(t)}={ \int_{M_{min}}^{M_{max}} \overline{n}(M) p(M,t,w) {\rm d}M \over \int_{M_{min}}^{M_{max}} \overline{n}(M) f(M,t){\rm d}M}$$ where $M_{min}$ is the mass of stars which are at two magnitudes below the turn off, $M_{max}$ the mass of stars which are at the turn off, $\overline{n}(M)$, the number of stars born with an initial mass $M$, and $p(M,t,w)$ the fraction of these stars which, at the age $t$, are on the MS and have surface velocities such that $\upsilon/\upsilon_\mathrm{crit} \ge w$, $f(M,t)$ is the fraction of stars of initial mass $M$ which are on the MS at the time $t$. In the following we explain how these different values are obtained from our rotating stellar models.
\begin{figure*}
  \resizebox{\hsize}{!}{\includegraphics{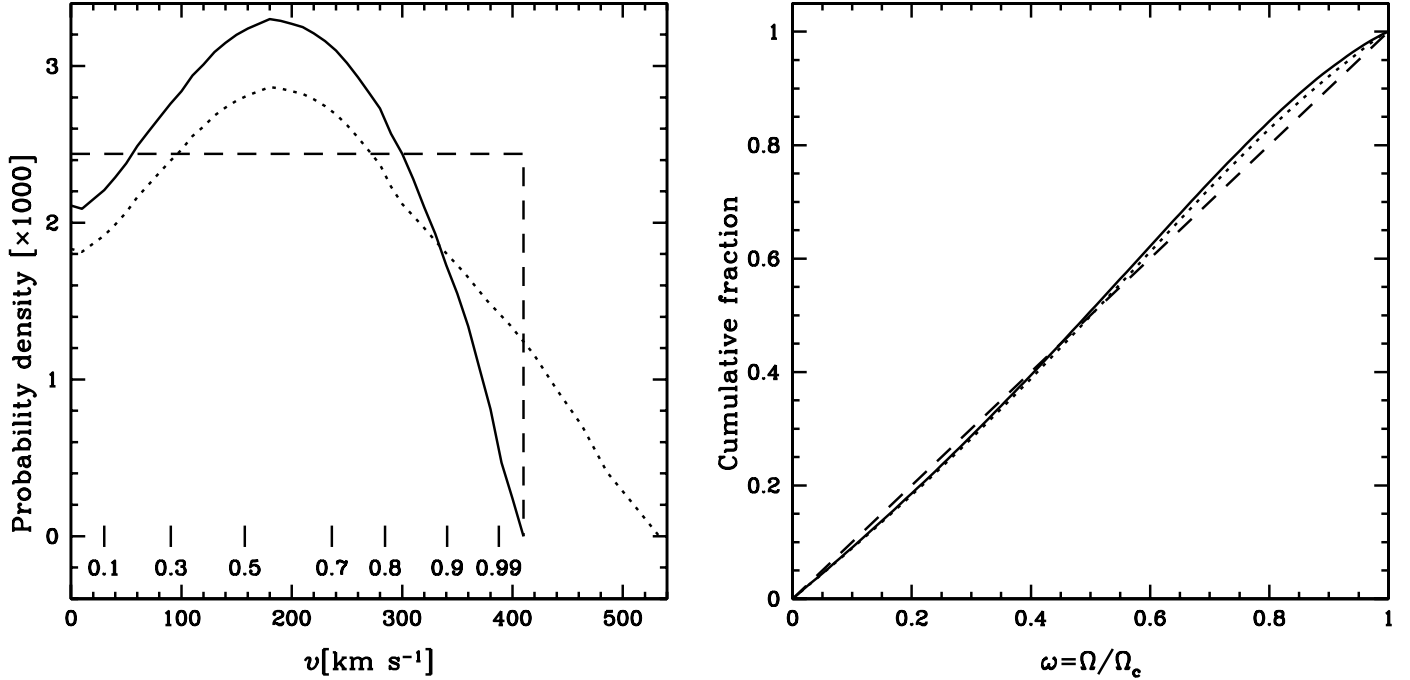}}
      \caption{\emph{Left panel:} distribution of the equatorial velocities as given by \citet{HG06a} (continuous line). The dotted line shows the case when it is assumed that above 300 km s$^{-1}$, the velocities are underestimated (see text). The dashed line shows a flat distribution. At the bottom of the figure, the correspondence between ZAMS values of $\Omega/\Omega_\mathrm{crit}$ and average velocities during the MS phase as deduced from our 9 $M_{\sun}$ at standard metallicity (Table \ref{tvmoy}) is indicated. \emph{Right panel:} cumulative distribution function of $\omega=\Omega/\Omega_\mathrm{crit}$ deduced from the data by \citet{HG06a} with the same assertions as the left panel.}
       \label{distrib}
  \end{figure*} 
\begin{itemize}
\item $\underline{M_\mathrm{max}}$: In the case of non-rotating stellar models, at a given metallicity we have a one-to-one relation between the initial mass and the MS lifetime. It suffices then to take 
the initial mass value of the star which has a MS lifetime equal to the considered age of the cluster in order to obtain the mass of stars at the turn off, {\it i.e.} $M_\mathrm{max}$.
When rotation is accounted for, the situation is more complicated, since the MS lifetimes increase with the initial velocities. Thus the maximum mass of a star which can still be on the MS at a given age corresponds to the mass of the rotating star, having the maximum initial rotation velocity (here $\Omega/\Omega_\mathrm{crit}=0.99$ on the ZAMS) and whose MS lifetime corresponds to the age considered. Of course in a given cluster there is little chance to have at turn off such a star due to the very small number of stars with such a high velocity. This effect will be taken into account through the initial velocity distribution function in $f(M,t)$ (see below).
\item ${\underline{M_\mathrm{min}}}$: To estimate the lower bound $M_\mathrm{min}$, we have taken about 70\% of the value of $M_\mathrm{max}$. We have indeed checked using theoretical isochrones \citep{meynet93} that for ages between 8 and 316 Myr, the typical mass of stars two visual magnitudes below the turn off is in that range.
\item $\underline{\overline{n}(M)}$: The number of stars born in the mass interval between $M_\mathrm{min}$ and $M_\mathrm{max}$ is proportional to $\int_{M_\mathrm{min}}^{M_\mathrm{max}} M^{-2.35}{\rm d}M$, when a Salpeter's slope of the initial mass function is considered.  
\item $\underline{f(M,t)}$: as explained above, stars with a mass $M_\mathrm{max}$ will still be in their MS phase, provided they started their evolution with $\Omega/\Omega_\mathrm{crit} \ge 0.99$. Stars of this mass but with slower initial rotation will have already evolved away from the MS. Thus the fraction of stars of mass $M_\mathrm{max}$ which are still on the MS at time $t$ is equal to
$$f(M_\mathrm{max},t)=\int_{\omega_\mathrm{min}(M,t)=0.99}^{\omega=1.00} d_p(\omega) \mathrm{d}\omega$$
where $\omega_\mathrm{min}(M,t)$ is the minimum value of $\omega$ for stars of initial mass $M$ to be still on the MS at the time $t$, and $d_p$ the probability density for a star to start its evolution with a given value $\omega$ (See Fig.~\ref{distrib}). For each mass and age, one can determine $\omega_\mathrm{min}(M,t)$, and similar integrals as the one shown above can be performed.
\item $\underline{p(M,t,w)}$: From figures similar to Fig.~\ref{vevol}, we can extract for each initial mass, metallicity and rotation, the time at which $\upsilon/\upsilon_\mathrm{crit}$ becomes superior to $w$ (note that once $\upsilon/\upsilon_\mathrm{crit} \ge w$, it remains superior to it for the rest of the MS phase\footnote{Note that we used the extrapolated lifetimes (marked with an asterisk in Table 3),
for the models whose computation was stopped before the completion of the MS phase.}, the only exception being the 20 $M_{\sun}$ at $Z=0.020$). From such data, one can deduce that for a star of 9 $M_{\sun}$ with $Z=0.020$ and an age equal to 25 Myr, the minimum value of $\Omega/\Omega_\mathrm{crit}$ required on the ZAMS to reach values of $\upsilon/\upsilon_\mathrm{crit}$ superior to 0.7 during the MS phase is 0.83.
Using an initial distribution of $\Omega/\Omega_\mathrm{crit}$ (see Fig.~\ref{distrib}), one can deduce the value of $p(M,t,w)$.
\end{itemize}

We need now to specify the function we use for the initial distribution of $\Omega/\Omega_\mathrm{crit}$. \citet{HG06a} provide from their observations of 496 OB-type stars a distribution of equatorial velocities corrected for the $\sin i$ effect. This distribution is shown in Fig.~\ref{distrib} (left).
Of course this distribution does not correspond to a distribution on the ZAMS, stars being picked up at different phases of their evolution during the MS phase. We can however reasonably consider that these equatorial velocities are representative of the values of the average velocities of stars in the MS phase.
We can then associate a value of $\omega$ on the ZAMS to each these values using the relation obtained from our 9 $M_{\sun}$ at standard metallicity between the averaged equatorial velocity during the MS phase and $\omega$ on the ZAMS.
Proper normalization have been applied to transform the probability densities for velocities into probability densities for $\omega$, \emph{i.e.} the normalization has been chosen so that 
$\int_0^{420}D_p(\upsilon) \mathrm{d}\upsilon=\int_0^1 d_p(\omega)\mathrm{d}\omega=1$,
where $D_p$ and $d_p$ are the probability densities for respectively $\upsilon$
(see  Fig.~\ref{distrib}, left panel) and $\omega$. The cumulative distribution of $\omega$ on the ZAMS we obtained in this way is shown in the right panel of Fig.~\ref{distrib}. 
\begin{figure*}
  \resizebox{\hsize}{!}{\includegraphics[angle=-90]{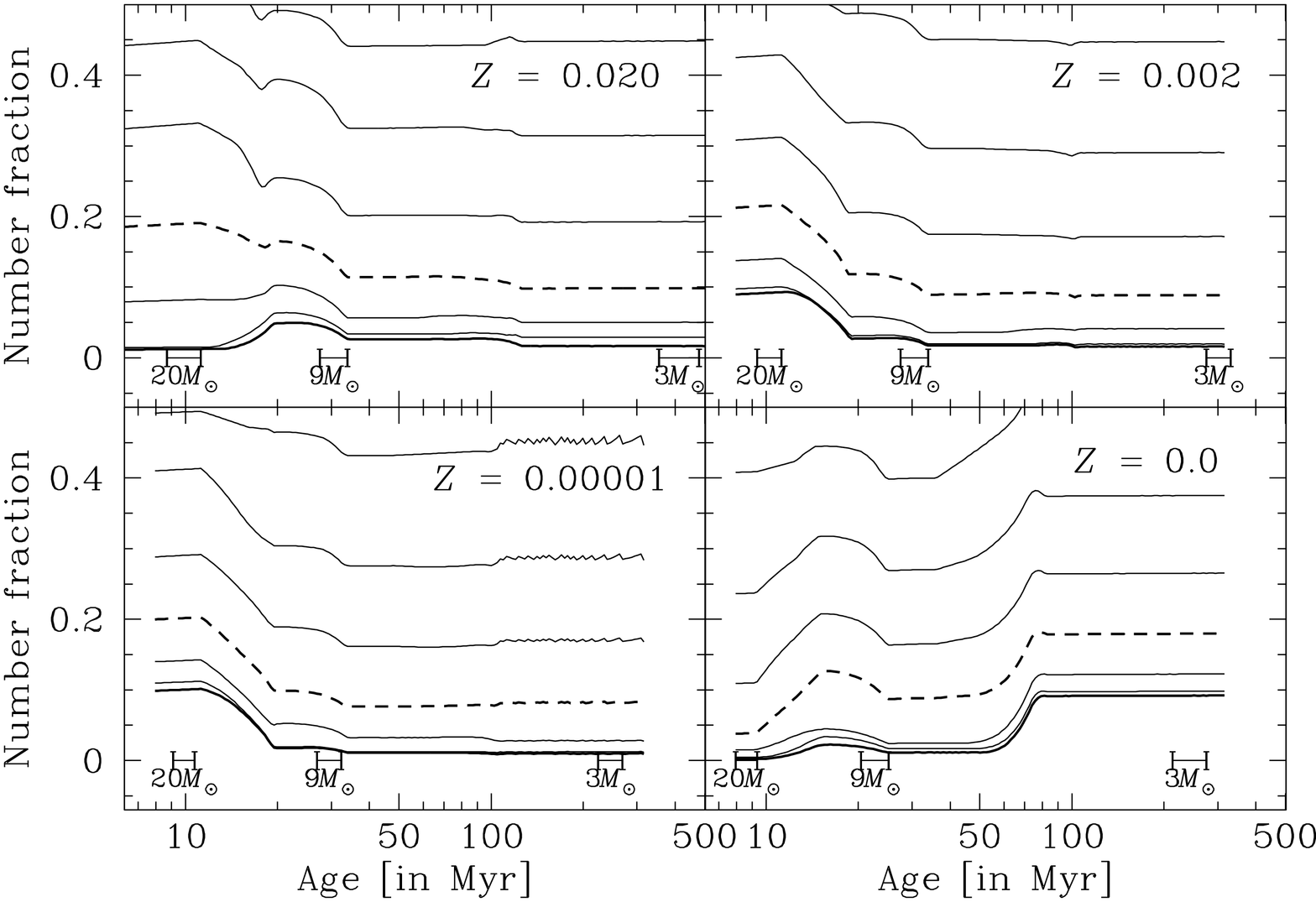}}
      \caption{Variation of the fraction of stars with $\upsilon/\upsilon_\mathrm{crit}$ superior or equal to 0.4, 0.5, 0.6 (three thin continuous lines from the top), 0.7 (thick dashed line), 0.8, 0.9 (two thin continuous lines) and at the critical limit (bottom thick solid line) in clusters of various ages and metallicities. At the bottom of each panel, the range of MS lifetimes for the 3, 9 and 20 $M_{\sun}$ stellar models are indicated.}
       \label{fraca}
  \end{figure*}

In Fig.~\ref{fraca} the variation of the number fraction of stars at the critical limit (continuous thick lines) in clusters of different ages is shown. We have plotted also the fraction of stars having $\upsilon/\upsilon_\mathrm{crit}$ values superior to given limits (see caption). 
Due to the coarse grid of mass we used here, we shall not attach too much importance to the details of these curves but will comment on their general behaviour. 
A few interesting features can be underlined:
\begin{itemize}
\item At standard metallicity, we note a bump around ages 20 - 25 Myr for the curves corresponding to the critical limit and for $\upsilon/\upsilon_\mathrm{crit} \ge 0.9$ and 0.8. This kind of behaviour can be explained in the following way: at younger ages than the bump, mass loss disfavours the reaching of the critical limit, at older ages than the bump, meridional currents are not efficient enough for transporting outwards the angular momentum (more dense envelopes in stars of lower initial masses).
\item For metallicities $Z=0.002$ and 0.00001, we note that the bump shifts to younger ages and reaches higher values. This is due to the smaller mass loss rates experienced by stars at low metallicity. This allows more massive stars, which have less dense envelopes and thus stronger meridional currents, to undergo a more rapid acceleration of their surface and to reach more easily the critical limit. 
\item We note that the curves are very similar at the metallicities $Z=0.002$ and 0.00001. Thus, from the point of view of the surface velocities, we would not expect any significant changes over this metallicity range.
\item For metal-free stars the curves are very different than at $Z=0.00001$ and 0.020: at ages younger than about 20 - 32 Myr, stars have more difficulty to reach the critical limit. This comes from the fact that the absence of metals make these stars very compact, with dense outer layers and very slow meridional circulation currents (note that would the mass range be extended towards higher masses, it is possible that this bump would appear at very young ages). At ages older than about 80 Myr, on the other hand, stars at the critical limit are much more numerous than at higher metallicities. This comes from the fact that stars with initial masses equal or below 3-5 $M_{\sun}$ contract for a great part of their MS phases (see the metal-free 3 $M_{\sun}$ stellar models) and thus can reach quite early the break-up limit.
\item  At $Z=0.020$, the fraction of stars at the critical limit is of the order of at most 5\% and can reach values of slightly more than 10\% at lower metallicities. The number fraction of stars with $\upsilon/\upsilon_\mathrm{crit} \ge 0.9$ tightly follows the curve of stars at the critical limit, indicating that when this value is reached, the surface velocity rapidly evolves towards the critical limit. The fraction of stars with $\upsilon/\upsilon_\mathrm{crit} \ge 0.7$ reaches values up to 20-25\% and this at an earlier age: around 10 Myr and below.
\end{itemize}

\section{Discussion of the results\label{discuss}}

\subsection{Sensitivity on the ZAMS model}

We recalled in Sect.~4.2 that our way of building our ZAMS models makes it impossible to
maintain a high value of $\upsilon/\upsilon_\mathrm{crit}$ from the ZAMS all along the beginning of the MS phase. The activation of the meridional circulation imposes a rapid slow down of the surface and thus, the star needs to somewhat evolve in the MS phase in order to approach again the critical limit. At this point, one can ask two questions: 1) would it be possible to build ZAMS models with higher angular momentum content, which would remain near the critical limit all along during the MS phase? 2) How would such behaviour, if possible, change the above results concerning the frequency of stars near the critical limit? 

The answer to the first question is yes. It would suffice to distribute differently the angular momentum or the angular velocity in the ZAMS model, for instance imposing on the ZAMS a non-uniform distribution of $\Omega$ allowing more angular momentum to be contained in the inner parts of the star. But this would not change the present results concerning the frequencies of stars near the critical limit. Indeed, these frequencies are computed on the basis of the observed frequencies of stars with given averaged rotational velocities during the MS phase. Stars with higher angular momentum than the most rapid rotators computed here will present higher values of the averaged rotational velocities during the MS and are thus already implicitly accounted for.

\subsection{Sensitivity on the velocity distribution}

Let us first note that the measure of the rotational velocity through the Doppler effect is not free from spurious effects. First this technique gives access only to $\upsilon \sin i$ where $i$ is the angle between the line of sight and the rotational axis. This quantity cannot be known except if some other observations as e.g. a rotational period obtained from variability induced by the presence of dark spot at the surface are available. But even in that case, the knowledge of $\upsilon$ requires the knowledge of the radius (more precisely of the equatorial radius). Moreover if the surface rotates differentially like the solar surface, the period thus obtained might only be partially representative of the surface rotation.
Second due to the von Zeipel theorem, the fast rotating regions of the star, i.e. the zones near the equator, are less luminous than the polar regions which have small linear velocities. This effect produces an underestimate of the rotational velocities. \citet{Town04} obtain the following behaviour between the FWHM of the HeI 4471 line for a B2-type star and the equatorial
velocity: the relation is nearly perfectly linear for $\sin i=1$ and $\upsilon/\upsilon_\mathrm{crit} < 0.7$ (for a standard metallicity 9 $M_{\sun}$ star, this corresponds to velocities inferior to 300 km s$^{-1}$ on the ZAMS). In that velocity regime, the FWHM measurements, due to the above effects,
underestimate the velocities by at most 3-4\%. For  $\upsilon/\upsilon_\mathrm{crit} > 0.7$, the relation is no longer linear, the FWHM increases much more slowly with the velocity, it even goes through a maximum, decreasing a little for velocities near the critical limit. In that case, the FWHM becomes a poor indicator of the surface velocity, a star rotating at the critical limit presenting the same line broadening as a star rotating at 80\% the critical velocity. Said in other words, this technique may underestimate the real velocity of the star by 20-25\% for very fast rotators (typically for $\upsilon/\upsilon_\mathrm{crit} > 0.75$, {\it i.e.} for velocities superior to about 340 km s$^{-1}$, for a 9$M_{\sun}$ at standard metallicity). 
To study this effect, we have done the following experiment: we have modified the velocity distribution given by \citet{HG06a} assuming that between 300 and 420 km s$^{-1}$ all the velocities $\upsilon$ should be enhanced by a  factor equal to 1+($\upsilon$-300)/120*0.3 (\emph{i.e}. a linear enhancement amounting at most to a factor 1.3 at 420 km/s). Using such a modified velocity distribution, shown by the dotted line in Fig.~\ref{distrib}, we obtain that the fraction of stars at the critical limit for an age equal to 25 Myr at $Z=0.020$ would be equal to 0.055 instead of 0.049 with the original velocity distribution of \citealt{HG06a} (continuous line in Fig.~\ref{distrib}). Thus this effect would not significantly change the results. 

In order to obtain a fraction of stars at the critical limit of about 10\% at $Z=0.020$, it is necessary
to considerably increase the fraction of stars born with $\Omega/\Omega_\mathrm{crit}$ superior to 0.6 on the ZAMS. Typically passing from the original velocity distribution (continuous line in Fig.~\ref{distrib})
to a uniform distribution (dashed line), one obtains that the fraction of stars with values of $\Omega/\Omega_\mathrm{crit}$ on the ZAMS superior to 0.6, 0.7, 0.8 and 0.9, passes from 37.5, 26.5, 15.5 and 6.5\% to 40, 30, 20 and 10\% respectively. Using such a uniform velocity distribution, one obtains that the fraction of stars at the critical limit equal 12\% at $Z=0.020$ and age equal to 25 Myr (instead of 5\% with the original velocity distribution).

\subsection{Sensitivity on mass loss and magnetic fields}

Mass loss, as we have already discussed above, plays a key role in preventing the high mass stars
to reach the critical limit. 
We showed that using two different laws for the mass loss rate, very different behaviours are obtained for the 60 $M_{\sun}$ models at standard metallicity. It underlines the importance of this ingredient in shaping the evolution of the surface velocity in this mass range.
For lower initial mass stars, typically for 20 $M_{\sun}$, mass loss becomes less and less important (at least in absence of any surface magnetic field).

We can wonder whether mass loss anisotropies would have an impact on the results obtained in this work (which assumed spherical winds). Let us recall that when a star is rotating at $\upsilon/\upsilon_\mathrm{crit} \ge 0.9$, winds of hot stars will become much more intense at the pole than at the equator. This effect would accelerate the evolution towards the critical limit. However, accounting for this effect would hardly change the evolution we computed here. Indeed, $\upsilon/\upsilon_\mathrm{crit}$ becomes superior to about 0.8-0.9 only when the surface velocity increases anyway in an already very rapid way (see Fig.~\ref{vevol}).

A magnetic field, following the Tayler-Spruit dynamo \citep{spruit02}, will tend to impose solid body rotation during the MS phase \citep{mm3}, thus making the approach of the critical limit easier (stronger coupling between the core and the envelope). We can expect that the inclusion of this effect would have an important impact on the present results, making the fraction of stars at the critical limit more important than in the present work.
It has however to be noted that at present the Tayler-Spruit dynamo is still much debated and some more work is needed before assessing whether this mechanism can be active in stars or not.

\section{Possible links with the Be stars\label{becomp}}
%
\begin{figure}
  \resizebox{\hsize}{!}{\includegraphics[angle=-90]{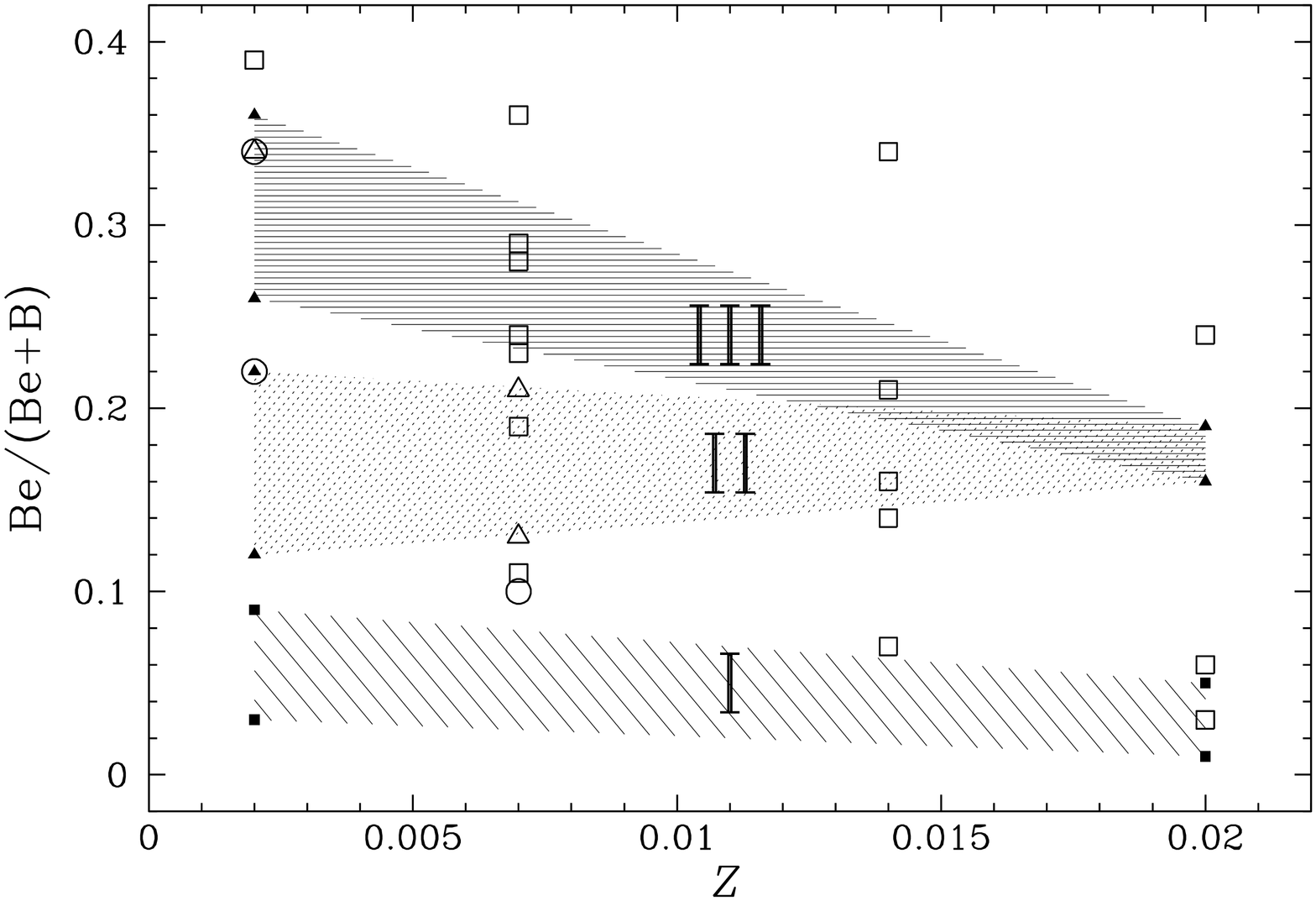}}
      \caption{The empty symbols correspond to the observed fraction of Be stars among B0-B3 type objects in clusters with ages ranging from 10 to 25 Myr. Data are taken from \citealt{WB06} (circles), \citealt{kel99} (triangles) and \citealt{MGM99} (squares). The filled symbols correspond to predictions of the given models for $Z=0.002$ and $Z=0.020$ and for ages equal to 10 and 25 Myr respectively (we estimated the fraction of Be in a two magnitude interval below the turn off, as did \citealt{MGM99}).
      The hatched zone labelled by I correspond to stars predicted at the critical limit. The hatched zone labelled by II shows the fractions of stars with $\upsilon/\upsilon_\mathrm{crit}\ge 0.7$. These two cases were
      obtained assuming a velocity distribution of \citet{HG06a}. The third hatched region, labelled III corresponds to stars with $\upsilon/\upsilon_\mathrm{crit}\ge 0.7$ and assuming that at low metallicity the
      initial distribution of rotation contains faster rotators (see text).}
       \label{compobs}
  \end{figure}
 
There are good indications that the Be star phenomenon is related to the fast rotation of the star. 
For instance \citet{Martayan} showed that the initial velocities of the Be stars are significantly higher than the initial velocities of the normal B stars, giving some support to the view that only stars with a sufficiently high initial velocity can go through a Be episode. Let us note that when material is ejected into a Keplerian disk  the star loses angular momentum, slows down and evolves away from the limit. Secular evolution will then bring back the star towards this limit. Probably the phases in between the loss of material are much longer than the phases during which the star loses material into the Keplerian disk. Thus most of the Be stars might be observed during these quiet phases when they are evolving back towards the limit.
It means that the rotation of observed Be stars might be somewhat lower that the rotation at which outbursts occur.
If the link between Be stars and fast rotation is now well accepted, it is however not clear what rotation rate is needed for a star to become a Be star, or why this process is restricted mainly to early B type stars. Is rotation the only key parameter governing the transformation of a star into a Be star? When does such a transformation occur?
In the present section we would like to provide some answers to the above questions.

Comparisons between number fractions of fast rotating stars as predicted by the present models (filled symbols and hatched regions) and as observed in clusters (empty symbols, see caption) is shown in Fig.~\ref{compobs}. We see first that if we consider only stars that are strictly at the critical limit (hatched zone I), the prediction falls well below the observed points, all the more so as there are also rapidly rotating stars that are not Be stars, or Be stars that are observed between outbursts. Thus according to the present stellar models, the number fraction of stars at the critical limit is not sufficient to account for the observed fraction of Be stars.

Let us mention that there are two proposed channels for the formation of Be stars: 1) the single star channel, where the critical limit is reached because of the natural evolution of a single rotating star; and 2) the binarity channel, with the spin-up of the member of a binary system through mass transfer from its companion. According to \citet{McSG05a,McSG05b}, about 75\% of the Be stars detected may have been spun-up by binary mass transfer, while most of the remaining Be stars were likely rapid rotators at birth. There are indeed some known binaries among the observed Be stars, but in most cases, no binarity has been yet clearly assessed. It may however be very difficult to detect, because of the long orbital period and the very small orbital motions, and in some cases, decades of careful measurements have been necessary before detecting a companion (see \citealt{Gies98} for $\phi$ Per, \citealt{Maintz05} for 59 Cyg, \citealt{Riv04} for FY CMa). According to \citet{neg07}, the binarity channel may be significant, but it is unlikely that most Be stars may form through this channel. From a theoretical point of view, \citet{Pols91} deduce that probably 40\% (at most 60\%) of the Be stars may form through mass transfer spin-up. Consequently, the binary channel alone could account for a (Be/(B+Be)) ratio of the order of 8-10\% if we take the (Be/(B+Be)) value found by \citet{Mart06}. In that last case, our prediction remains thus too low even when discounting the binarity channel contribution.

In the frame of the single star scenario, if we assume that the Be phenomenon may occur already at $\upsilon/\upsilon_\mathrm{crit}\ge 0.7$ (see hatched zone II), then the situation improves somewhat, however the theoretical results do not show any increase of the fraction of Be stars with decreasing metallicity, as some observations indicate \citep{MGM99,WB06}. It might be that part of this trend is due to an observational bias. For instance \citet{McSG05b} found that Be stars were generally found among the brighter B-stars in their survey of southern sky Galactic clusters. A similar trend was found by \citet{Kel01} in h and $\chi$ Per. Since metal poor clusters are more distant, observations will pick up mainly the brighter end of the B stars sample where the fraction of Be stars is higher. However, it is not clear that such a bias will have a great impact when clusters of similar ages are compared: provided the observed samples are complete and the count is performed in the same absolute magnitude interval, then the observed number ratios  at different metallicities can be compared. To reproduce the observed trend with the metallicity obtained by \citet{MGM99} and more recently by \citet{WB06} with the present models, we have to assume that the fraction of fast rotators at birth is higher in metal poor regions than in metal rich one. As a numerical example, supposing that the initial distribution of rotation at $Z=0.002$ is given by a uniform distribution (dashed line in Fig.~\ref{distrib}), we then obtain the hatched zone III, which agrees reasonably well with the observations. From the discussion above, we deduce, that to reproduce the observational number fraction of Be stars, first we have to relax the hypothesis that Be stars are strictly at the critical limit: the Be phenomenon should occur already for $\upsilon/\upsilon_\mathrm{crit} \ga 0.7$; second, the distribution of initial values of $\Omega/\Omega_\mathrm{crit}$ on the ZAMS at low metallicity should contain a higher fraction of fast rotating stars.

Let us discuss the two above conclusions.
The first conclusion would tell us that, although fast rotation plays an important role in transforming the star into a Be star, it is not sufficient alone. Above some value of $\upsilon/\upsilon_\mathrm{crit}$ some other processes, like for instance pulsations, must help in launching material into a disk.

Now what about the second conclusion: a higher fraction of fast rotators at low Z?
\citet{kell04} measured the rotational velocity of 100 MS early B-type stars in clusters of the LMC and compared the results with observations of early B-type stars in clusters of the solar neighbourhood. He obtains that the LMC stars are faster rotators than the galactic ones: the mean values of $\upsilon\sin i$ is 116 km s$^{-1}$ for the galactic stars and 146 km s$^{-1}$ for the LMC stars. Part of the difference can be due to the difference in mass loss between galactic and LMC stars, but part might also be due to differences in the initial distribution of the velocities.
\citet{Penny04}, on the other hand, find no difference between the velocities of O-type stars in the Magellanic Clouds and in the Galaxy. Thus the picture is at present not clear and probably 
new surveys, as e.g. the VLT-Flames survey will provide very soon new insights on that question. 
At the moment we still consider the question as opened. Present results would however clearly favour a higher fraction of stars born with high values of $\Omega/\Omega_\mathrm{crit}$ at low metallicities.

A point which the models have to explain is when the Be phenomenon does appear during the MS phase. Unfortunately no clear picture emerges from the observations: some authors find the highest fraction of Be stars among the spectral types B0-B2, appearing most often at ages between 13-25 Myr \citep{FT00}. Recently \citet{FN06} who determined the fundamental parameters of the Be stars which would be observable by COROT found that the main part of their stellar sample falls in the second half of the MS lifetime. This would favour a view according to which the Be star phenomenon appears preferentially at the end of the MS phase. Other authors \citep[as \emph{e.g.}][]{ZB97} conclude that the Be phenomenon is not associated with a particular evolutionary phase and can occur during the whole MS phase, at least for some range of initial masses \citep[see][]{ZF05}. With the adopted distribution of initial velocities, present models predict that most near critical velocity stars will be found at evolved stages during the MS phase and would thus provide an explanation for the observation of \citet{FT00}. From a physical point of view, we think that the Be star phenomenon may be limited to a given spectral type range for the following reasons: for the earlier spectral types than the favoured one, mass loss prevents the star from approaching the critical limit and for the later spectral types, the meridional currents are not efficient enough for accelerating the envelope. Models would predict a shift towards earlier spectral type when the metallicity decreases due to the decrease of mass loss with the metallicity (extending the range towards earlier spectral type) and the decrease of the meridional velocity in denser layers (increasing the minimum initial mass for meridional currents being efficient enough to accelerate the surface). Of course, beyond these physical reasons are the effects of a possible variation of the distribution of initial rotation as a function of metallicity, which would somewhat blur the above picture. And now if we relax the condition for becoming a Be star to $\upsilon/\upsilon_\mathrm{crit} \ga 0.7$, we allow this phenomenon to appear earlier in the life of a star.

The present models predict the approach of high values of $\upsilon/\upsilon_\mathrm{crit}$ (0.8-0.99) only when the star is already evolved into the MS phase. As mentioned above, the use of a different internal profile of $\Omega$ in our ZAMS models would allow us to explore still more extreme cases of rotation, where the star may reach the critical limit, or at least remain near to it, from the beginning of the MS phase.
However these stars should be very rare. The necessity for the stars to be somewhat evolved in order
to reach high surface velocities implies that these stars should present at their surface N-enrichments.
These surface enrichments might however be very small for stars in the lower part of the mass range possibly evolving into a Be star phase (see the cases of our 3 $M_{\sun}$ stellar model at $Z=0.020$ in Fig.~\ref{ncp}). 
Observations of N surface enrichments for Be stars are still scarce, probably because of the difficulty of determining abundances from greatly widened absorption lines. \citet{FZ04} obtain N/C ratios for four Be stars which are clearly higher than what is observed at the surface of slowly rotating B-type stars. They obtain (N/C)/(N/C)$_\mathrm{ini}$ of the order of a factor 2, which would well correspond to what is expected for near critically rotating 9 $M_{\sun}$ stars at at $Z=0.020$ (see middle panel of Fig.~\ref{ncp}).

\section{Conclusions \label{conclu}}
From the above computations we can deduce that
\begin{enumerate}
\item As long as the mass loss rates are not too important, meridional currents, by advecting angular momentum from the inner regions to the outer layers, bring the $\upsilon/\upsilon_\mathrm{crit}$ ratio close to 1 during the MS phase.
\item Since the velocity of the meridional currents in the outer layers scales with the inverse of the density, the process becomes more efficient for stars of higher initial mass and/or higher initial metallicity.
\item At high metallicity however, mass loss becomes more and more important and can prevent the stars from reaching the break--up limit.
\item In stellar clusters, one expects that the fraction of stars having $\upsilon/\upsilon_\mathrm{crit} \ga 0.8$ becomes maximum for ages between 20-32 Myr at $Z=0.020$. This range of ages shifts to older ages at higher metallicities and to younger ages at lower metallicities. Such a shift results from the dependence on the metallicity and the mass of the intensity of the mass loss rates and of the velocities of the meridional currents. If we relax the condition and look at stars with $\upsilon/\upsilon_\mathrm{crit} \ga 0.7$, the maximum number will be found in clusters of ages around 10 Myr and below, and this at all non-zero metallicities.
\item Be stars might be the natural outcome of stars with an initial rotational velocity in the upper tail of the initial velocity distribution. To explain their frequency from the present rotating stellar models, minimum values of $\upsilon/\upsilon_\mathrm{crit} \ga 0.7$ have to be reached during the MS for the Be star phenomenon to occur. 
\item Depending on when this limit is reached, one expects more or less high surface enrichments. If
the limit is reached very early during the MS phase, no enrichment is expected, while if this limit is reached at the end of the MS phase, high N/C and N/O ratios are expected. For Be stars originating from small initial mass stars (typically from 3 $M_{\sun}$ at standard metallicity) one expects no or small surface enrichments.
\item To reproduce the higher fractions of Be stars at low metallicity, we must assume that in metal poor regions, a higher number of stars are born with high values of $\Omega/\Omega_\mathrm{crit}$.
\end{enumerate}
In a future work, it will be interesting to explore the effects of a magnetic field. As already mentioned above, this will probably change significantly the results presented here, allowing the stars to approach more easily the critical limit. It will be interesting to see if, in that case, it would still be necessary to invoke a higher proportion of fast rotators at low $Z$. 

\begin{acknowledgements}
We are very grateful to the referee Douglas R. Gies for his careful reading. His valuable comments have helped us improving the presentation of this paper.
\end{acknowledgements}
\bibliographystyle{aa}
\bibliography{EMMB}

\end{document}